\documentclass[aps,prd,nofootinbib,twocolumn,floatfix,superscriptaddress,preprintnumbers,showpacs]{revtex4}

\usepackage[english]{babel}
\usepackage{graphicx}
\usepackage{color}
\usepackage{amsmath,amsfonts}
\usepackage{subfigure}

\DeclareMathOperator{\Tr}{Tr}
\DeclareMathOperator{\tr}{tr}

\DeclareMathOperator{\Pexp}{Pexp}

\newcommand{\diff}{\mbox{d}}
\newcommand{\llangle}{\left\langle}
\newcommand{\rrangle}{\right\rangle}

\newcommand{\beq}{\begin{eqnarray}}
\newcommand{\eeq}{\end{eqnarray}}

\newcommand{\D}{\mathcal{D}}

\newcommand{\os}{\overline{\s}}

\renewcommand{\tr}{\mbox{Tr}}

\newcommand{\m}{\mu}

\newcommand{\s}{\sigma}

\renewcommand{\D}{\Delta}

\newcommand{\oh}{\frac{1}{2}}

\newcommand{\non}{\nonumber}

\newcommand{\ra}{\rightarrow}

\begin{document}
\bibliographystyle{h-physrev5}

\title{$k$-string tensions and the $1/N$ expansion} 

\author{J. Greensite}

\affiliation{Physics and Astronomy Dept., San Francisco State
University, San Francisco, CA~94132, USA}
\affiliation{The Niels Bohr Institute, DK-2100 Copenhagen \O, Denmark}

\author{B. Lucini}
\affiliation{School of Physical Sciences, Swansea University, Singleton Park, Swansea SA2 8PP, UK}

\author{A. Patella}
\affiliation{CERN, Physics Department, 1211 Geneva 23, Switzerland}
 
\date{\today}

\begin{abstract}

We address the question of whether the large-$N$  expansion in pure SU($N$) gauge theories requires that
$k$-string tensions must have a power series expansion in $1/N^2$, as in the sine law, or whether $1/N$ 
contributions are also allowable, as in Casimir scaling. 
We find that $k$-string tensions may, in fact, have $1/N$ corrections, and consistency with the large-$N$ expansion
in the open string sector depends crucially on an exact cancellation, which we will prove, among terms involving odd powers of $1/N$ in particular combinations of Wilson loops. It is shown how these cancellations are
fulfilled, and consistency with the large-$N$ expansion achieved, in a concrete example, namely, strong coupling lattice gauge theory with the heat-kernel action.  This is a model which has both a $1/N^2$ expansion and Casimir scaling of the $k$-string tensions. Analysis of the closed string channel in this model confirms our conclusions, and provides further insights into the large-$N$ dependence of energy eigenstates and eigenvalues. 

\end{abstract}

\preprint{CERN-PH-TH/2011-016}

\maketitle

\section{Introduction}
\label{sec:intro}

One of the signatures of confinement in SU($N$) Yang-Mills theory (cf.\ \cite{Greensite:2003bk} for a recent review) is the asymptotic linearity of the static quark potential as a function of quark-antiquark separation.  The generally accepted interpretation is that a chromoelectric flux tube (or \textit{open string}) forms between the two static charges, and the slope of the potential in the asymptotic regime is the string tension.  In general we may consider the static quark charge to be in some representation $R$ of the gauge group, with the antiquark in the conjugate $\overline{R}$ representation.   At small and intermediate distances the static quark potential depends on $R$, but asymptotically, due to color-screening by gluons, the force between quark and antiquark depends only on the $N$-ality $k \in [0,N-1]$ of the SU($N$) representation, with $k < N/2$ and $N -k$ being equivalent
by charge conjugation.  If $k$ is non-zero, then the asymptotic string tension $\s_k$ is also non-zero, and depends only on $k$.  One may also consider flux tubes wrapping around a spatial compact direction (\textit{closed strings}), which are not attached to any static charge. These states can also be characterized by their $N$-ality, and their asymptotic string tension is the same as the one in the open string channel.  This is a manifestation of the open-closed string duality in Yang-Mills theory.

    The dependence of $\sigma_k$ on $k$ and $N$ is an important probe into the dynamics
of confinement, since different confinement mechanisms suggest different dependencies. 
For this reason, various lattice studies have been undertaken to determine the ratios
$\s_k/\s$ of the $k$-string tensions to the fundamental
representation string tension.  A question of particular interest is whether the first correction to the ratio at infinite
$N$ is of order $1/N$ or $1/N^2$.\footnote{This is the leading correction for fixed $k$.  Another case of interest is the 
$N\ra \infty$ limit with the ratio $k/N$ held fixed, but this case will not be considered here.}   While early lattice results~\cite{Wingate:2000bb,Lucini:2000qp,Lucini:2001nv,DelDebbio:2001kz,DelDebbio:2001sj,Lucini:2004my} were inconclusive on this point,
more recent simulations~\cite{Bringoltz:2008nd} provide evidence that the finite-$N$ corrections
begin at order $1/N$.  On the face of it, this result seems to go against what we might expect from a large-$N$ expansion, which for pure $SU(N)$ gauge theories is an expansion in powers of $1/N^2$.  This objection has been made in some detail in Refs.\  \cite{Armoni:2003ji,Armoni:2003nz}, where it is explicitly argued, on the basis of the large-$N$
expansion, that the asymptotic string tension can only have large-$N$ corrections
that can be expressed in a power series of $1/N^2$.   If that large-$N$ reasoning is correct, then it can be used to argue in favor of the sine law (see the next section)
and to rule out Casimir scaling of $k$-string tensions.\footnote{It was also argued that the order $1/N$
correction can only enter the amplitudes in correlators~\cite{Armoni:2006ri}.}   Yet the Monte Carlo data seem to contradict this seemingly straightforward conclusion.

    In this article we ask whether large-$N$ arguments, and specifically those of Refs.\  \cite{Armoni:2003ji,Armoni:2003nz}, unavoidably lead to $1/N^2$ rather than $1/N$ corrections for $k$-string ratios (which would
rule out Casimir scaling in particular).  We will argue below that the answer is no.   In the picture which emerges
we will find that the energy levels of the  open string, with static charges $R$ and $\overline{R}$ at its ends, contain in general odd powers in $1/N$ for a generic irreducible representation $R$.  The asymptotic string tension, associated with the ground state of the open string, also contains odd powers of $1/N$.  Despite appearances, this result is not in conflict with the large-$N$ expansion.  In particular, the large-$N$ expansion of the expectation value of a product of (fundamental representation) Wilson loops is a power series in $1/N^2$, in conformity with the standard large-$N$ counting.  

We also find that the large-$N$ expansion of the energy levels of a closed $k$-string, wrapping around a spatial compact dimension, contains only powers of $1/N^2$ at every fixed compactification radius.  Nevertheless, the ground state of the closed $k$-string has an asymptotic string tension which coincides with the one computed in the open string channel, that in general contains odd powers of $1/N$.  This apparent contradiction will be resolved in Sect.~\ref{sect:closed}. 

    Most of our explicit calculations are carried out in the framework
of the lattice strong coupling expansion with the heat-kernel
action.   This is a theory in which the large-$N$ expansion, and
Casimir scaling, are known to coexist; what we are after is to
understand \emph{how} they can coexist.   Of course the
strong coupling lattice theory differs from the 
continuum theory both quantitatively and qualitatively in various
ways, e.g.\ there is no L\"uscher term in the static potential.  But the
strongly-coupled theory also has many realistic features expected in
the continuum theory, and among these are  confinement by a linear
potential, chiral symmetry breaking, color screening, and the related
decay of metastable $k$-strings into stable strings. We will argue
that some of the features we find are not limited to the
strong coupling expansion, as they are based on general properties of
correlators and of the large-$N$ limit.

   We should note here that some subtleties that could invalidate the conclusions
of~\cite{Armoni:2003ji,Armoni:2003nz} have already been pointed out
in~\cite{KorthalsAltes:2005ph}. Analogies and differences between our
arguments and those presented in~\cite{KorthalsAltes:2005ph} will also be discussed below. 

The rest of our paper is organized as follows.
In Sect.~\ref{sect:review} we briefly review some of the relevant literature on
$k$-strings and their large-$N$ limit, pointing out why a better
theoretical understanding is needed, and anticipating our findings.
Sect.~\ref{sect:duality} recalls some salient facts about the determination
of string tensions, in the open and closed string channels, from Wilson loops and 
Polyakov loop correlators respectively. Some general facts about
large-$N$ counting for Polyakov and Wilson loops are discussed in
Sect.~\ref{sect:rcconjg}, where the concept of RC-conjugate
representations will also be introduced. In Sect.~\ref{sect:wl} we
elaborate on the large $N$ behavior of
Wilson loops in  tensor product representations and prove
the absence of odd-power corrections in $1/N$ in Wilson loops in these
reducible representations. In Sect.~\ref{sect:hk} we
discuss in detail the case of lattice
SU($N$) gauge theory with the heat-kernel action, and demonstrate
that a cancellation mechanism is at work, among the odd powers in $1/N$ contributing
to product loops, in the leading-order contributions of the lattice strong coupling expansion. 
This shows explicitly why $1/N$ corrections in $k$-string tension ratios
cannot be ruled out in general, on the basis of
large-$N$ arguments alone. A complementary analysis performed in terms
of Polyakov loop correlators provides a different perspective on why
possible $1/N$ corrections in string tensions are in general allowed (Sect.~\ref{sect:closed}).
After a further discussion of the presence or absence of energy degeneracies in the large-$N$ limit
(Sect.~\ref{sect:discussion}), we report our conclusions in Sect.~\ref{sect:conclusions}. Finally, in the
appendices we provide proofs of some group-theoretic statements made in the text.

\section{$k$-strings and large-$N$}
\label{sect:review}
$k$-string tensions may provide important insights into the problem of confinement, since
their dependence on $k$ and on the fundamental string tension is determined
by the effective degrees of freedom responsible for
confinement. For a setup with sources in a representation $R$, the asymptotic
string tension is expected to depend only on the $N$-ality, i.e.\ on
the integer $k$ describing the phase of the source after performing a
$\mathbb{Z}_N$ transformation, and not on the representation
itself (by charge conjugation, $N$-ality $k<  N/2$ and $N - k$ are equivalent). This is easily understood if we start from closed strings and
advocate the open-closed string duality, which will be discussed in some
detail in Sect.~\ref{sect:duality} to establish the equality of the
string tensions in both setups. In the absence of external sources and if a spatial direction is compactified, center transformations (i.e. gauge transformations periodic up to a center element) are Hamiltonian symmetries of the Yang-Mills system. Hence the charge of the Polyakov loop
under the center (which coincides with the $N$-ality of the representation in which the Polyakov loop is considered) classifies eigenstates of the Hamiltonian in the closed string
channel. As a consequence, the $N$-ality labels possible independent string tensions. By open-closed string duality, the $N$-ality can be used to label string tensions also in the open string channel.

Different effective models of confinement, and various beyond the
Standard Model frameworks, make differing predictions for the ratios of these string tensions.
In particular, the so-called sine law for $k$-string tensions 
\beq
\frac{\sigma_k}{\sigma} = \frac{\sin\left(\frac{\pi
      k}{N}\right)}{\sin\left(\frac{\pi}{N} \right)} \ ,
\eeq
where $\s=\sigma_1$ is the fundamental string tension, tends to arise in
supersymmetric models~\cite{Douglas:1995nw,Armoni:2003ji}, in MQCD \cite{Hanany:1997hr}, and in certain AdS/CFT-inspired models \cite{Herzog:2001fq}.\footnote{This does not mean that the sine law is necessarily exact in these models.  Higher-order corrections to the sine law, in the case of softly broken ${\cal N}=2$ supersymmetric 
SU($N$) gauge theory, are derived in ref.\ \cite{Auzzi:2002dn}.}
Other frameworks
favor Casimir scaling instead.   The term ``Casimir scaling'' \cite{DelDebbio:1995gc}
originally referred to string tensions at intermediate distances, prior
to color screening by gluons, where it is supposed that $\s_R \approx (C_R/C_F) \s$.
Here $C_R$ denotes the quadratic Casimir of representation $R$, and $C_F$ is the
quadratic Casimir of the fundamental defining representation.   This behavior is
exact in 1+1 dimensions, and it also arises, at intermediate distance scales,
from the conjectured dimensional-reduction form of the Yang-Mills vacuum 
wavefunctional \cite{Greensite:2007ij,Karabali:2009rg}, as well
as from the stochastic vacuum picture \cite{DiGiacomo:2000va}.  
In fact, there have been a number of studies which report that string tensions of metastable flux tubes,
associated with static sources in various color representations prior to color screening, 
are related by Casimir scaling \cite{Ambjorn:1984mb,Deldar:1999vi,Bali:2000un}.

Representations which have the minimal groundstate energy
for a given $N$-ality cannot be screened by gluons, and one may then suppose
that the string tensions $\s_k$ of these representations obey the Casimir scaling rule not only at
intermediate distance scales, but also
asymptotically.  In that case, the prediction for $k$-strings is 
\beq
{\s_k \over \s} = {C_{ka} \over C_F} = {k(N-k) \over N-1} \ ,
\eeq
where $C_{ka}$ is the quadratic Casimir of the representation corresponding to the
antisymmetrized product of $k$ fundamental defining representations.   
Casimir scaling of $k$-string tensions is found in certain
supersymmetric models~\cite{Auzzi:2008ep,Kneipp:2003ue}\footnote{In
  Ref.~\cite{Kneipp:2007fg}, the role of the particular choice of the
  superpotential in determining $k$-string tension ratios has been investigated.},
and, as the leading term in an appropriate expansion, in gauge-adjoint
Higgs models in 2+1 dimensions \cite{Antonov:2003tz,Antonov:2004es}.
A number of scenarios other than Casimir scaling and the sine law
(such as string counting, in which $\s_k = k\s$, or the bag model
inspired ratio of square-root Casimirs) are already excluded by
lattice simulations~\cite{Wingate:2000bb,Lucini:2000qp,Lucini:2001nv,DelDebbio:2001kz,DelDebbio:2001sj,Lucini:2004my}. 

A striking difference between the sine law and Casimir
scaling is that while sine law $k$-string tensions have an expansion
in only even powers of $1/N$, the same expansion for Casimir scaling
$k$-string tensions contains both even and odd powers, with the first
correction to the large-$N$ limit being of order $1/N$.  This difference was
first noted by Strassler in ref.\ \cite{Strassler:1997ny}, who emphasized the
relation of the sine law to mechanisms found in supersymmetric gauge theories, as well as 
the importance of calculating $k$-string tensions via lattice simulations.
Although in SU($N$) gauge
theory there is no compelling reason to expect exact sine law or exact
Casimir scaling behaviour, it is important to at least understand
whether the $1/N$ expansion of $k$-string tensions contains only even
powers, or instead can involve both even and odd powers of the
expansion parameter.  This issue has been discussed
in~\cite{Armoni:2003ji,Armoni:2003nz,Armoni:2006ri}, and it was argued in
\cite{Armoni:2003nz}  that odd powers in $1/N$ (and therefore Casimir scaling in particular) are
excluded from appearing in $k$-string tension ratios, on general grounds of the large-$N$ expansion. 
However, as we have already noted, lattice Monte Carlo simulations have also weighed in on this issue. 
While the question is not settled in 3+1 dimensions, 
in 2+1 dimensions the most precise lattice simulations~\cite{Bringoltz:2008nd} provide
evidence that the first correction to $k$-string tensions in the large $N$ limit is of order
$1/N$ rather than $1/N^2$.

   It is important to understand how the conflict between the lattice data, and the 
(apparent) requirements of the large-$N$ expansion, can be resolved.
An explanation which suggests that lattice simulations are measuring the intermediate, rather than
the asymptotic string tensions~\cite{Armoni:2003nz} would not really resolve the discrepancy.
It is, of course, possible that there are non-perturbative effects in the 
intermediate region which evade the usual large-$N$ counting.  However, if large-$N$ arguments 
can break down at intermediate distances, then it is not clear to us why one should be persuaded 
by large-$N$ arguments regarding the asymptotic behavior.

   On the other hand, if the lattice data on $k$-string tensions is in fact probing the
asymptotic region, and if the large-$N$ expansion is not misleading us, then there
is a second possibility worth exploring: it could be that there is a cancellation mechanism at work.
We take as given the fact that the expectation values of certain Wilson and Polyakov loops, in some (generally reducible) representations, can be expanded in a power series in $1/N^2$ only. But the key point is whether this fact
\emph{necessarily} implies that the same type of expansion holds for the associated energy
levels, and in particular for the static quark potential.  What we suggest is that in cases where energy levels 
with corrections of order $1/N$ appear in correlators that must have a $1/N^2$ expansion, there is an
exact cancellation among the odd powers of $1/N$.  

  The possibility of a cancellation among odd powers of $1/N$, which was not considered
in~\cite{Armoni:2003nz}, has been studied in the context of an
effective quantum mechanical Hamiltonian
in~\cite{KorthalsAltes:2005ph}. In that work, the authors argue that
at infinite $N$ there is a degeneracy in the spectrum among closed
string states of the same $N$-ality.  Open strings are characterized
by the group representation of the static sources at their ends (see
e.g.~\cite{Luscher:2002qv,Armoni:2009zq}), and a degeneracy among
these states, at infinite $N$, is understandable. 
There is no reason, however, that closed string energy eigenstates of
a given $N$-ality should correspond precisely to a particular group
representation, and if a degeneracy among such states emerges in the
large-$N$ limit, it raises a new question:  \textit{Degeneracies are usually associated
  with symmetries.  Is there some new, and until now unrecognized,
  global symmetry which emerges in the large-$N$ limit?  Or,
  alternatively, do non-degenerate states in the closed string sector
  remain non-degenerate at $N = \infty$?  How would the supposed
  cancellation mechanism work in the latter case?} 

The purpose of this work is to investigate these open issues. We will find, as already noted in the Introduction, that
\begin{enumerate}
\item In the Wilson loop ({\em open string channel}) approach, the
  vacuum expectation value of the Wilson loop 
is representation-dependent, although, as expected, the asymptotic
string tension is only $N$-ality dependent. In the general case the
string tension can have $O(1/N)$ corrections, and consistency with the
large-$N$ expansion 
depends crucially on an exact cancellation, which we will prove, among
terms involving odd powers of $1/N$ in particular combinations of
Wilson loops.  These combinations are the sum of two Wilson loops in
representations $R,R'$, whose Young tableaux are related by swapping
rows and columns. Crucial to this mechanism is a well-defined
finite-$N$ splitting of the states leading to the cancellation, which
become degenerate only at exactly $N = \infty$.  The cancellation
condition is respected by Casimir scaling, which therefore cannot be
excluded by large-$N$ arguments.  
\item In the spacelike Polyakov loop ({\em closed string channel})
formalism, the cancellation of the odd powers of $1/N$ works a
little differently, and we find that the energy eigenvalues can be
expanded in even powers of $1/N$ only. However, the string tension
can still have corrections of $O(1/N)$, and in fact this quantity
agrees with the string tension in the open channel, as it should.
We will explain how energies can have only O($1/N^2$) corrections,
yet the asymptotic string tension has $1/N$ corrections. 
\end{enumerate}

\section{Open and closed strings}
\label{sect:duality}
String tensions can be determined from either Wilson loops, or Polyakov loop correlators.  The string tensions in 
either approach are related, for reasons which we will briefly recall.

In Euclidean space, let $U^C$ denote a Wilson loop holonomy  
\beq
\label{eq:holonomy}
    U^C   =  \Pexp \left(i\oint_C A^\m \diff x_\m  \right)  \ , 
\eeq
with
\beq
     W^C_R =   \langle \Tr_R U^C \rangle
\eeq
the expectation value of the trace of the holonomy in the group representation $R$; i.e.\ 
$\Tr_R U \equiv \chi_R(U)$ where $\chi_R$ is the group character.\footnote{Note also that $\Tr_R \mathbb{I} = d_R$, 
where $d_R$ dimension of the representation.}
If not otherwise stated, $C$ is chosen as a rectangle with edges of length $r$ and $\tau$ lying respectively along a spatial direction $\hat{r}$ and along the
temporal direction. In the Hamiltonian formalism, the Wilson loop can be seen as the correlator of a quark-antiquark pair joined by a thin flux line sitting at time zero and relative distance $r$ and the conjugated of the same operator at time $\tau$. The states propagating in the correlator are precisely the states of the \textit{open string} with static charges in the representations $R$ and $\overline{R}$ at its ends.\footnote{We will always assume 
that $R$ has non-trivial $N$-ality.}    Decomposing in energy eigenstates one gets:
\beq
\label{corrr}
W_R = \sum_n \left| \alpha_{n}(r) \right|^2 e^{ - V_{n}(r) \tau} \ .
\eeq
The ground state energy $V_0(r)$ is the static potential between the static charges.

On a torus with spatial radius $r$, the Wilson loop holonomy in Eq.~\eqref{eq:holonomy} can be taken along the
compact direction. Its trace defines the Polyakov line:{\footnote{ Polyakov loops in spacelike directions are also known as \emph{torelon} operators, although we will not use that term here.}  
\beq
P_R(x,y,t) = \Tr_R U^C \ ,
\eeq
where the compactified direction has been assumed to be $z$ and the coordinates $(x,y,t)$ identify the contour
$C$. As in the case of a Wilson loop, the zero-momentum correlator of a Polyakov loop and a conjugate Polyakov loop separated by a time lapse $\tau$ can be expanded in energy eigenstates:
\beq
\label{eq:polycorr}
\langle P_R^{\dag} (0,0,0) \sum_{x,y} P_R (x,y,\tau) \rangle = \sum_n \left| \beta_{n}(r) \right|^2 e^{ - E_{n}(r) \tau} \ ,
\eeq
where the states entering in the right hand side now are {\em closed string} states wrapping around the torus.

We can reinterpret $r$ as time and $\tau$ as spatial distance, which is allowed, since in Euclidean space all the directions are equivalent. The correlator of Polyakov loops becomes (going to the Hamiltonian formalism) the partition function of two static quarks at distance $\tau$ and temperature $1/r$. Expanding in energy eigenstates gives
\beq
\label{eq:polyopen}
\langle P_R^{\dag} (x,y,t) P_R (x,y,t+\tau) \rangle = \sum_n m_n e^{ - V_{n}(\tau) r} \ .
\eeq
Now it is open string states which propagate, hence the energies $V_n$ are the same as the ones appearing in Wilson loops in Eq.~\eqref{corrr}. The fact that the same quantity, namely the correlators of Polyakov loops, can be expanded in open and closed string states is known as open-closed string duality and gives constraints on the functional form of the energy levels (see e.g.~\cite{Luscher:2004ib} and references therein). The most direct consequence of the open-closed string duality is the identity of the string tensions in the two channels:
\begin{equation}
\lim_{r \to \infty} \frac{V_0(r)}{r} = \lim_{r \to \infty} \frac{E_0(r)}{r} = \sigma_k \ .
\end{equation}
One important difference between the open and the closed string
channels is the dependence on the representation. In the open channel,
the states that propagate can be characterized by the representation
of the static sources at the ends of the string. 
In contrast, since there are no static sources in the closed string
channel, the spectrum as a whole can only be characterized by
$N$-ality. It is possible that a particular energy eigenstate of a given
$N$-ality, at certain distance scales, may project largely onto a
subspace corresponding to Polyakov lines in a definite group
representation~\cite{Bringoltz:2008nd,Athenodorou:2008cj}, but there
is no reason that the energy eigenstates should be contained entirely
in that subspace.\footnote{The issue is addressed quantitatively, in
  the context of the strong-coupling expansion, in
  section~\ref{sect:closed}.} Thus, in Polyakov line correlators, we 
expect that all energy eigenstates with the $N$-ality of the 
representation of the Polyakov lines will propagate between 
the Polyakov loop and its conjugate.  In the Wilson loop case, only
eigenstates with appropriate static sources at the 
endpoints will propagate.  Of course, despite this difference, the asymptotic string
tension in both the open and the closed string channel is expected to
be dependent on the $N$-ality only.  In the 
closed string channel this is true because all states of the $N$-ality
of the Polyakov lines will propagate, and the result is dominated by
the lowest energy state. In the open string channel, the fact that the
asymptotic string tension depends only on the $N$-ality of the
representation is guaranteed by color screening.\footnote{Color
  screening also guarantees a dependence of open string energy
  eigenvalues on the group representation of the static sources even
  at large color source separations, since the screening process
  introduces an $r$-independent contribution to the energy. Even in the
  absence of color screening there will be $r$-independent self-energy
  contributions (at least at weak-couplings, due to perturbative
  one-gluon exchange) which depend on the color representation of the
  static sources at the endpoints.} 

Although the open and the closed string channels are expected to
provide the same answer for the string tension, there may be important
differences in the way those answers are determined in each
channel. For this reason, we shall perform our analysis in both
cases. 

\section{Correlators in the large-$N$ theory}
\label{sect:rcconjg}
Consider the operator $O_R(r,\tau)$, which can be either the
zero-momentum product of two Polyakov loops,
\begin{equation}
O_R(r,\tau) =  P_R^{\dag} (0,0,0) \sum_{x,y} P_R (x,y,\tau) \ ,
\end{equation}
 or the (conveniently normalized) trace of the Wilson loop holonomy, 
\begin{equation}
O_R(r,\tau) = \frac{1}{N^k} \Tr_R \left(\Pexp \left(i \oint_C A^\m \diff x_\m  \right)
\right) \ ,
\end{equation}
in the representation $R$. Let $R$ be a representation of 
SU($N$) of $N$-ality $k$, whose corresponding Young tableau
is formed from $k < N$ boxes.  In both cases the vacuum
expectation value (VEV) of $O_R$ will have a power
expansion in $1/N$ that starts with a term of order one. In general,
for a generic reducible representation, the leading correction is of
order $1/N$. It is convenient to consider linear combinations of 
$\langle O_R(r,\tau)\rangle$ in appropriate representations $R$, such that their expansion in $1/N$ contain only powers of $1/N^2$.

\begin{figure}
\begin{center}
\includegraphics[height=3.0cm]{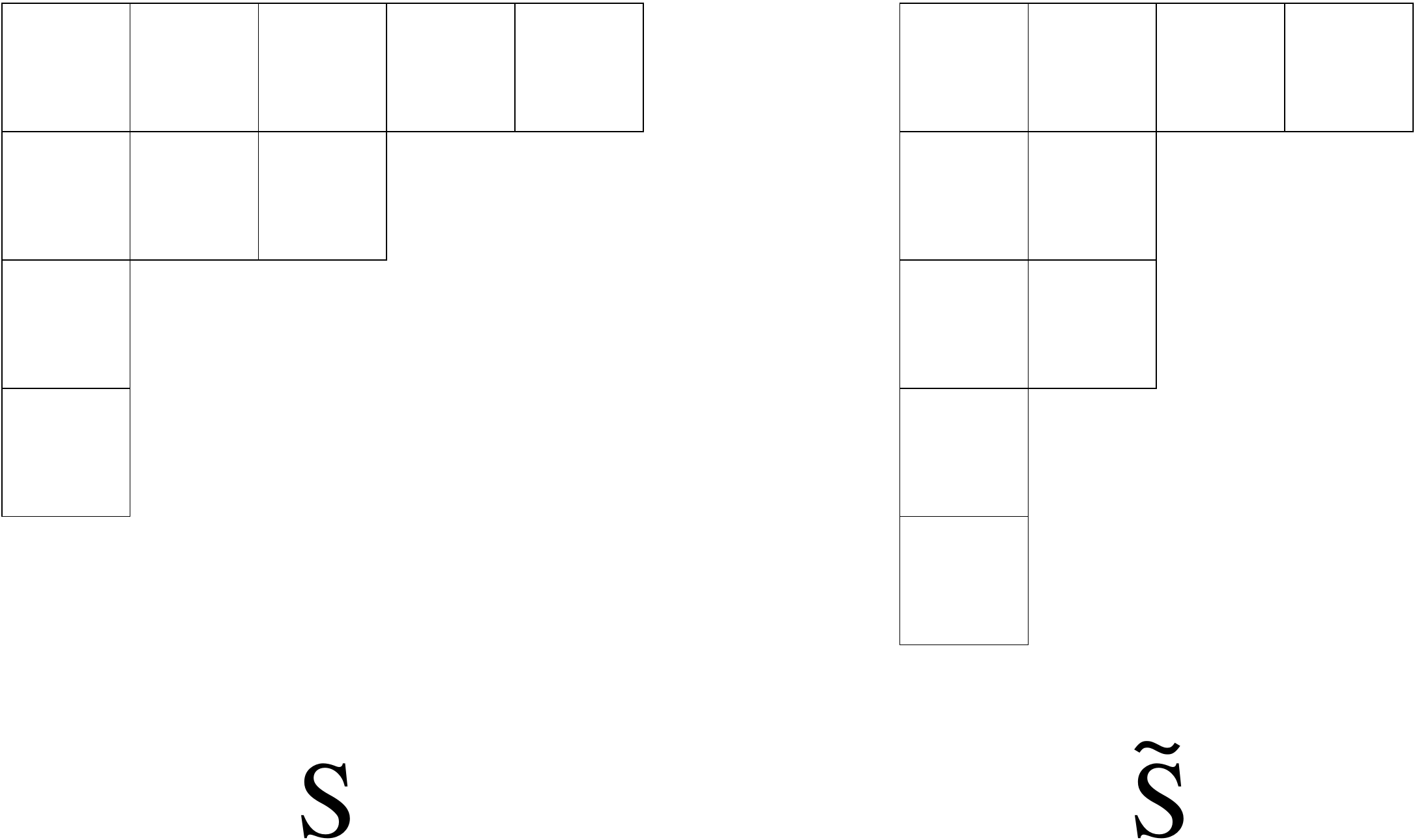}    
\end{center}
\caption{An example of an RC-conjugate pair of Young tableaux $(S,\widetilde{S})$.}
\label{fig:young_conj}
\end{figure}

Consider a Young tableau $S$
corresponding to some irreducible representation $R \equiv R_S$, and imagine constructing a new tableau
by interchanging the rows and columns of the original tableau.  That is, the first (topmost) row
of the original tableau becomes the first (leftmost) column of the new tableau, the second row of
the original tableau becomes the second column of the new tableau, and so on, as illustrated in 
Fig.~\ref{fig:young_conj}.  The group representation $\widetilde{R} \equiv
R_{\widetilde{S}}$ associated with the new tableau $\widetilde{S}$ will 
be called the row-column conjugate
(``RC-conjugate'') of representation $R$.  In particular, the fully symmetric and fully antisymmetric
representations of $N$-ality $k$ are RC-conjugate representations.  If $R$ and $\widetilde{R}$ are the same 
representation, then the representations are RC self-conjugate.  

In Appendix~\ref{appendix:characters} we prove that if $R$ and
$\widetilde{R}$ are RC-conjugate representatios, the expectation
values $\langle O_R(r,\tau) \rangle$ and $\langle
O_{\widetilde{R}} (r,\tau) \rangle$ are related to each other via the
transformation  $N \to -N$.\footnote{Transformation properties under $N
  \to$ $- N$ were also considered by Lohmeyer, Neuberger, and
  Wettig~\cite{Lohmayer:2009aw} in connection with their study of the
  density of eigenvalues of Wilson loops in two dimensions. Although
  the context is quite different, some of their expressions are in
  form reminiscent of relationships that we derive here.}
Hence, the combination
\begin{equation}
\label{eq:rcconjcorr}
\llangle O_R(r,\tau) \rrangle + \llangle O_{\widetilde{R}} (r,\tau) \rrangle 
\end{equation}
can be expressed in a power series in $1/N^2$.  The VEVs appearing
in~\eqref{eq:rcconjcorr} can both be expanded in the appropriate Hamiltonian eigenstates:
\begin{gather}
\label{eq:wilson_S}
\llangle O_{R} 
\rrangle = \sum_n A_{n}(r) e^{ - E_{n}(r) \tau} \ , \\
\label{eq:wilson_Stilde}
\llangle O_{\widetilde{R}} 
\rrangle = \sum_n \widetilde{A}_{n}(r) e^{ - \widetilde{E}_{n}(r) \tau} \ .
\end{gather}
Although their sum must contain only powers of $1/N^2$, the large-$N$ expansion of
$\langle O_{R} \rangle$ and  $\llangle O_{\widetilde{R}}  \rrangle$ may each contain odd powers
of $1/N$ that cancel in the sum.  In particular, the fact that $N \ra -N$ takes 
$\langle O_{R} \rangle \ra \llangle O_{\widetilde{R}}  \rrangle$ implies a corresponding relationship
between the $A_n,E_n$, and the $\widetilde{A}_{n},\widetilde{E}_{n}$.
Viewing the expectation value $\langle O_{R} \rangle $ as a function of $N$, we consider taking
$N \to -N$, i.e.
\begin{flalign}
\llangle O_{R} 
\rrangle = \sum_n A_n(r,N) e^{ - E_n(r,N) \tau} 
\to \nonumber \\
\qquad \to \sum_n A_n(r,-N) e^{ - E_n(r,-N) \tau}
\ .
\end{flalign}
As we have already remarked, the result must be equal to the expectation value $\llangle O_{\widetilde{R}}  \rrangle $ for every $r$ and $\tau$. Since the expansion in exponentials is unique, it follows that the states can be  arranged so that the two conditions
\begin{gather}
\begin{split}
\widetilde{A}_n(r,N) = A_n(r,-N) \ , \\
\widetilde{E}_n(r,N) = E_n(r,-N) \ .
\end{split}
\label{eq:evenodd}
\end{gather}
are satisfied. States are then automatically organized so that amplitudes and energies indexed by $n$, in the two channels $R$ and $\widetilde{R}$, go to the same large-$N$ limit. 

To make the cancellation of odd powers of $1/N$ a little more explicit, it is useful to  separate the even and odd powers of $N$ by introducing the notation
\begin{flalign}
\begin{split}
&A_n^{\pm}(r,N) = \left( A_n (r,N) \pm A_n(r,-N) \right)/2 \ , \\ 
&E_n^{\pm}(r,N) = \left( E_n (r,N) \pm E_n(r,-N) \right)/2 \ . \\
\end{split}
\label{pm}
\end{flalign}
The sum of the two expectation values in eqs.~\eqref{eq:wilson_S} and~\eqref{eq:wilson_Stilde} then becomes
\begin{flalign}
& \llangle O_{R}  \rrangle + \llangle O_{\widetilde{R}} \rrangle = \nonumber \\
& = 2 \sum_n A_n^+(r) e^{-E_n^+(r)\tau} \cosh \left( E_n^-(r) \tau \right) + \nonumber \\
& \quad - 2 \sum_n A_n^-(r) e^{-E_n^+(r)\tau} \sinh \left( E_n^-(r) \tau \right) \ .
\end{flalign}
The cancellation of odd powers of $1/N$ occurs because the exponentials combine into $\cosh \left( E_n^-(r) \tau \right)$ and $A_n^-(r) \sinh \left( E_n^-(r) \tau \right)$, which contain only even powers of $1/N$.

Now suppose that  $E_n^-(r) = 0$.  This would imply, as a consequence of the second equation in~\eqref{eq:evenodd}, that the same spectrum of states contributes to RC-conjugate expectation values 
$\llangle O_{R}  \rrangle$ and  $\llangle O_{\widetilde{R}} \rrangle$. While this is the expected case for the Polyakov loop correlators, for which energy levels should depend only on the $N$-ality, the spectra of open strings (contributing to Wilson loops) should have some dependence on the representation of the static sources at the endpoints, and therefore on the representation of the loop.  This suggests that $E_n^-(r) \ne 0$ for open strings, in which case there must be odd power corrections in $1/N$ in the energy levels associated with Wilson loops.

Before proceeding further we will need to consider $-$ and dispose of $-$ the following line of reasoning:
The combination~\eqref{eq:rcconjcorr} has an expansion in powers of
$1/N^2$. When inserting the expansions~\eqref{eq:wilson_S}~and~\eqref{eq:wilson_Stilde},
this combination becomes a sum of terms, each of which falls off
exponentially with $\tau$.  There must be one term with the slowest
falloff, let us say $\exp[-E_0(r) \tau]$, and this term must dominate the
series as $\tau \ra \infty$.  It would seem to follow, then, that $E_0(r)$ must also have an expansion in powers of $1/N^2$.   This argument is fallacious, and to understand where
the fallacy lies it is instructive to consider the power series expansion of $\cosh(x)$.  Does the series contain
both even and odd powers of $x$, or only even powers?  The answer, of course, is that there are only even powers in the expansion. On the other hand, one could try to make the argument that for very large positive $x$, $\cosh(x)$ is very accurately approximated by $\oh e^x$, and therefore the function must have a series expansion in both even and odd powers of $x$! The argument is obviously wrong, but it illustrates a relevant fact:  To demonstrate that the power series expansion contains only even powers of $x$, it is necessary to keep both the growing exponential  \emph{and} the damped exponential terms, despite the fact that the damped exponential $\oh e^{-x}$ is negligible at large $x$. The odd powers of $x$ in the damped exponential exactly
cancel the odd powers of $x$ in the growing exponential, resulting in a power series with only even powers of $x$.

    We will refer to this analogy as the \textit{cosh argument}, and pursue it a little bit further.  Suppose we now consider the logarithm of $\cosh(x)$.   For small $x$, this function clearly has an expansion in only even powers of $x$,
i.e.\
\beq
     \log \cosh(x) = \oh x^2 - {1\over 12} x^4 + {1\over 45} x^6 - {17 \over 2520} x^8  + ...
\eeq
On the other hand, at $x \gg 1$, and $x \ll -1$
\beq
      \log \cosh(x) \approx |x|
\eeq
The right hand side is an even function of $x$, as it must be, yet for large positive or negative values it behaves,
in contrast to the small-$x$ expansion, like an odd power of $x$.  There is no contradiction between this large
$|x|$ behavior, and the fact that small-$x$ expansion contains only even powers of $x$.
 
    The cosh argument applies to the cases we are considering in two quite different limits, and it is important to understand that these limits do not commute.   In the first limit we keep 
$r,\tau$ fixed, and take $N$ very large.  In this limit, the sum of RC-conjugate loops~\eqref{eq:rcconjcorr} has a series expansion in powers of $1/N^2$, given that odd powers of $1/N$ cancel out when the two expectation values in eqs.~\eqref{eq:wilson_S} and~\eqref{eq:wilson_Stilde} are summed together.  This is the limit we would use to check that, e.g., Casimir scaling is consistent with the standard large-$N$ expansion.  The other, quite different limit is to keep $N$ fixed, and take $r$ or 
$\tau$ very large.  In this limit the expression~\eqref{eq:rcconjcorr}, viewed as a function of $N$, must still be invariant under $N \ra -N$, but the logarithm may involve factors of $1/|N|$, rather than
only even integer powers of $1/N$.   We will return to this point in Sect.~\ref{sect:closed}.  In either case there is no
requirement that the large-$N$ expansion of $E_0(r)$, or the energies of the other excited states, be restricted to only even powers of $1/N$.

\section{Wilson loops at large-$N$}
\label{sect:wl}
Let us begin our explicit analysis of the order of the correction with
$k$-strings in the open string channel. We will start by specializing
the analysis of the previous section to Wilson loops. 

The simplest examples of irreducible representations are the symmetric (2s) and antisymmetric (2a) representations 
of $N$-ality 2, whose Young tableaux are illustrated in the top line of Fig.~\ref{young}.  We have the identities
\beq
\frac{\llangle \Tr_{2a} U  \rrangle}{N^2} &=& \frac{1}{2N^2} \left\{ \llangle \left(\Tr U \right)^2 \rrangle - \llangle \Tr
  U^2 \rrangle\right\}  \ ,
\non \\ 
\frac{\llangle \Tr_{2s} U  \rrangle}{N^2} &=& \frac{1}{2N^2}\left\{ \llangle \left(\Tr U \right)^2 \rrangle + \llangle \Tr
  U^2 \rrangle\right\}  \ .
\label{astrace}
\eeq
where our convention is that $\Tr U$ with no subscript on $\Tr$ denotes the trace in the fundamental defining 
representation.  The usual large-$N$ counting arguments tell us that $\llangle (\Tr U )^2 \rrangle/N^2$ contains only even powers of $1/N$, while $\llangle \Tr U^2 \rrangle/N^2$ contains only odd powers of $1/N$.
Both $\llangle \Tr_{2a} U  \rrangle/N^2$ and $\llangle \Tr_{2s} U
\rrangle/N^2$ have a large-$N$ expansion which begins at $O(1)$, with
the leading correction of order $1/N$ rather than $1/N^2$. Moreover, they are related to each
other via the substitution $N \to -N$. This is expected, since the two
representations form an RC-conjugate pair.

The large-$N$ expansion also tells us that
\beq
          {1\over N^2}  \llangle \left(\Tr U \right)^2 \rrangle =    
               {1\over N^2}  \llangle \left(\Tr U \right) \rrangle^2 + O(1/N^2) \ ,
\label{fact}
\eeq 
which is an example of the large-$N$ factorization property.\footnote{In the $N\ra \infty$ limit, the expectation value of a product of gauge-invariant operators equals the product of their expectation values. This is known as ``factorization''.}  
This property can be derived from Feynman diagrammatic considerations, but it also holds in strong coupling lattice gauge theory (see, e.g., Ref.\ \cite{Makeenko:2002uj}).
Group theory dictates how to decompose the reducible product representation
into a sum of irreducible representations, and in this case
\beq
          {1\over N^2}  \llangle \left(\Tr U \right)^2 \rrangle =   {1\over N^2}\Bigl( \llangle \Tr_{2a} U  \rrangle 
             + \llangle \Tr_{2s} U  \rrangle \Bigr) \ .
\eeq
We see that the $1/N$ corrections to the expectation values of loops in the irreducible representations exactly 
cancel, so that the loop in the direct product representation has an expansion in powers of $1/N^2$ only.
Once again, this is in agreement with our general discussion of
RC-conjugate representations in Sect.~\ref{sect:rcconjg}.

In more generality, for a direct comparison
with~\cite{Armoni:2003ji,Armoni:2003nz},
let us consider Wilson loops in a tensor product of $k$ fundamental representations,
decomposed into a sum of  Wilson loops in irreducible representations
$R^i_k$ of $N$-ality $k$ ($k \le N/2$).   It is convenient to normalize these
operators by dividing by a factor of $N^k$. Let $g_{R_i^k}$
be the multiplicity of the representation $R_i^k$ in the decomposition of the tensor product
representation. Then
\beq
\label{corr}
{1\over N^k} \llangle \Tr_{\Box \otimes \Box \dots \otimes \Box} U  \rrangle &=& 
{1\over N^k}\llangle \left( \Tr U \right)^k \rrangle
\non \\
&=&{1\over N^k} \sum_i g_{R_i^k} \llangle \Tr_{R_i^k} U \rrangle \ ,
\eeq
where, in the standard Young tableaux notation, we have indicated
with $\Box$ the fundamental defining representation, and $\Box \otimes \Box
\dots \otimes \Box$ denotes the tensor product of $k$ fundamental representations.
As in the $k=2$ case,  one can use large-$N$ arguments to show that the Wilson loop in
the tensor product representation, divided by $N^k$, has a large-$N$ expansion in powers of
$1/N^2$.  This is equivalent to saying that, viewed as a function of $N$, the expectation value of the tensor
product is invariant under $N \to -N$. However, as we have seen
explicitly for the $k=2$ case, 
the same is not necessarily true for the irreducible representations $R_i^k$, in which odd 
powers in $1/N$ can appear.    

In the
decomposition of the tensor product of $k$ fundamental representations into a sum of
irreducible representations, RC-conjugate representations
enter with the same multiplicity $g_R = g_{\widetilde{R}}$.  Eq.~\eqref{corr} can be rewritten as
\beq
\label{corr_conj}
{1\over N^k}\llangle \left( \Tr U \right)^k \rrangle
= {1\over N^k} \sum_i g_i^k \left( \llangle \Tr_{R_i^k} U
  \rrangle + \llangle \Tr_{\widetilde{R}_i^k} U \rrangle\right) \ ,
\eeq
with the sum running over RC-conjugate pairs, and 
\beq 
g_i^k = \left\{ \begin{array}{cl}
                g_{R_i^k} /2 & \mbox{if $R_i^k$ is RC self-conjugate} \\
                g_{R_i^k}=g_{\widetilde{R}_i^k} & \mbox{otherwise}  \end{array} \right. \ .
\eeq
As we show in Appendix~\ref{appendix:characters}, the odd powers of
$1/N$ in VEVs of Wilson loops appear with opposite sign in
RC-conjugate representations. Hence the absence of odd powers in
$1/N$ in the left hand side of Eq.~\eqref{corr_conj} is a
consequence of the absence of odd powers in $1/N$ in each addendum
on the right hand side of~\eqref{corr_conj}.  

We will now turn to strong coupling lattice gauge theory, where we
will see the existence of precisely this pairwise cancellation
mechanism among the leading strong coupling contributions. In
addition, this example will show the existence of $1/N$ corrections in
the string tension, which appear as a consequence of
the advocated representation dependence of the Wilson loops.

\section{A lattice strong coupling example}
\label{sect:hk}

    In Sect.~\ref{sect:rcconjg} we derived some general conditions~\eqref{eq:evenodd} on amplitudes and energies, 
which are required for consistency with the large-$N$  expansion.  These conditions are general enough to allow corrections to the energy which include odd powers of $1/N$, as would be the case if $k$-string tensions followed the
Casimir scaling rule.  We will now investigate a calculable model which has both Casimir scaling for the string tensions,
and also a $1/N^2$ expansion.   This is simply the well-studied
example of strongly coupled lattice gauge theory, either in the Hamiltonian formulation, or in the Euclidean formulation
with a heat-kernel action.  Since the model has both Casimir scaling and a $1/N$ expansion, the cancellation
of odd powers of $1/N$ that we have discussed above ought to hold.  We will see that it does hold, at least for the
leading strong coupling contributions, and that in fact there is a pairwise cancellation of odd powers of $1/N$ among Wilson loops in RC-conjugate representations, as advertised.

   Let us begin with the Kogut-Susskind Hamiltonian 
\begin{flalign}
&H = {g^2 \over 2a} \sum_{l} E^a_l E^a_l + {1\over 2 g^2 a} \sum_p \mbox{Tr}[2 - U(p) - U^\dagger(p)] \ , \nonumber \\
\end{flalign}
where the sums are over links $l$ and spatial plaquettes $p$, respectively.   At very strong couplings
$g^2 \gg 1$, the leading approximation is obtained by simply neglecting the second sum over plaquettes,
and keeping only the first term.  Then for an on-axis static quark-antiquark pair at sites $\bf 0$ and
${\bf L}=L {\bf \hat{e}_i}$, with the quark in the lowest dimensional representation $R$ of $N$-ality $k$, the leading approximation to the
energy eigenstate is
\beq
           \Psi_{\overline{q}q} = \overline{q}({\bf 0}) \left(\prod_{n=0}^{L-1}  
                   U^{(R)}_i({\bf 0} + n{\bf \hat{e}_i}) \right) q({\bf L})  \Psi_0 \ ,
\eeq
where $\Psi_0$ is the vacuum state ($\approx 1$ to lowest order), and $U^{(R)}_i$ is the lattice link variable in color group representation $R$.  The energy of the quark-antiquark state, to leading order in $1/g^2$, is
\beq
            E_{\overline{q}q}(L) = g^2 C_R L \ ,
\eeq
and the corresponding string tensions $\sigma_R= g^2 C_R$ (in lattice units) obey the Casimir scaling rule.

   Since Casimir scaling at leading order does not depend on the potential term, let us consider a more general class of lattice strong coupling Hamiltonians  
\beq
          H = g^2 \sum_{l} E^a_l E^a_l +  \sum_p V[U(p)] \ ,
\eeq
and determine $V$ by requiring that $e^{-Ha}$ is the transfer matrix of some Euclidean lattice gauge theory on a hypercubic lattice, with an isotropic lattice 
action (i.e.\ the time direction is the same as all other directions).   This requirement, as shown by Menotti and Onofri \cite{Menotti:1981ry}, determines the potential term
to be
\beq
             e^{-aV[U(p)]} \propto \sum_R d_R e^{-g^2 C_R/2} \chi_R[U(p)] \ ,
\eeq
where the sum runs over all group representations $R$, and $\chi_R[U]=\Tr_R[U]$ is the group character.
The Euclidean lattice gauge theory which gives rise to this transfer matrix  has a lattice action
\beq
            e^{-S} = \prod_p \sum_{R_p} d_{R_p} e^{-g^2 C_{R_p}/2} \chi_{R_p}[U(p)] \ ,
\label{charex}
\eeq
where the index $p$ now runs over all plaquettes on the hypercubic lattice, and the sum is over all group representations.  This is known as the heat-kernel action, and it was introduced originally by 
Polyakov \cite{Polyakov:1978vu} and Susskind \cite{Susskind:1979up} in their pioneering treatments of the deconfinement phase transition (see also
Drouffe \cite{Drouffe:1978py}).    The strong coupling diagrams
for, e.g., Wilson loops and plaquette-plaquette correlators are identical in form for the heat-kernel and Wilson actions, which have the same large-N counting.  Of course the numerical value of each strong coupling diagram is different for the two actions, since the character expansion analogous to~\eqref{charex} for, e.g., the SU(2) Wilson action,  involves modified Bessel functions, rather than factors of $\exp[-{1\over 2} g^2 C_{R_p}]$.  As $g^2 \rightarrow 0$, the Wilson action converges to the heat-kernel action.

     Now we turn to Eq.~\eqref{corr}.  The leading contributions to the right-hand side of this equation are readily evaluated at strong couplings:
\begin{flalign}
&{1\over N^k} \sum_i g_{R_i^k} \llangle \Tr_{R_i^k} U \rrangle \nonumber \\
& = {1\over N^k} \sum_i g_{R_i^k} d_{R_i^k} \exp[-\s_{R_i^k} A] + {\rm np}_1  \ , 
\label{leading}
\end{flalign}
where 
\beq
          \s_R = {C_R \over C_F} \s = C_R {2N \over N^2-1} \s
\label{cas}
\eeq
is the Casimir scaling string tension of representation $R$, $d_R$ is the dimension of the representation, and np${}_1$ 
is the contribution from higher-order terms in the strong coupling expansion, which include non-planar terms beginning at order $1/N^2$.   Since $\s_R$ contains all powers,
even and odd, of $1/N$, one might expect that an expansion of the rhs of Eq.~\eqref{corr} in a power series in $1/N$ would also contain both even and odd powers of $1/N$.  If so, that would contradict the factorization property of the large-$N$ expansion, which, for the lhs of Eq.~\eqref{corr}, asserts that 
\beq
           {1\over N^k} \langle (\tr U)^k\rangle =  {1\over N^k} \langle \tr U \rangle^k + {\rm np}_2  \ ,
\label{condition}
\eeq
where np${}_2$ contains only even powers of $1/N$, starting with $1/N^2$.  In fact, individual contributions to the sum in~\eqref{leading} do contain terms of order
$1/N$.  We will now show explicitly, for the cases $k=2,3,4$, that when all terms in the sum are included, odd powers of $1/N$ disappear.  The cancellation, as we will see, occurs among RC-conjugate representations, and at leading order
in the strong coupling expansion the conditions~\eqref{eq:evenodd} have to be satisfied by the dimensionalities 
$d_R/N^k$, which are proportional to the $A_n$ coefficients, and the Casimir ratios $C_R/C_F$, which are proportional to the energies $V_n$.

\begin{figure}
\begin{center}
\includegraphics[height=6.0cm]{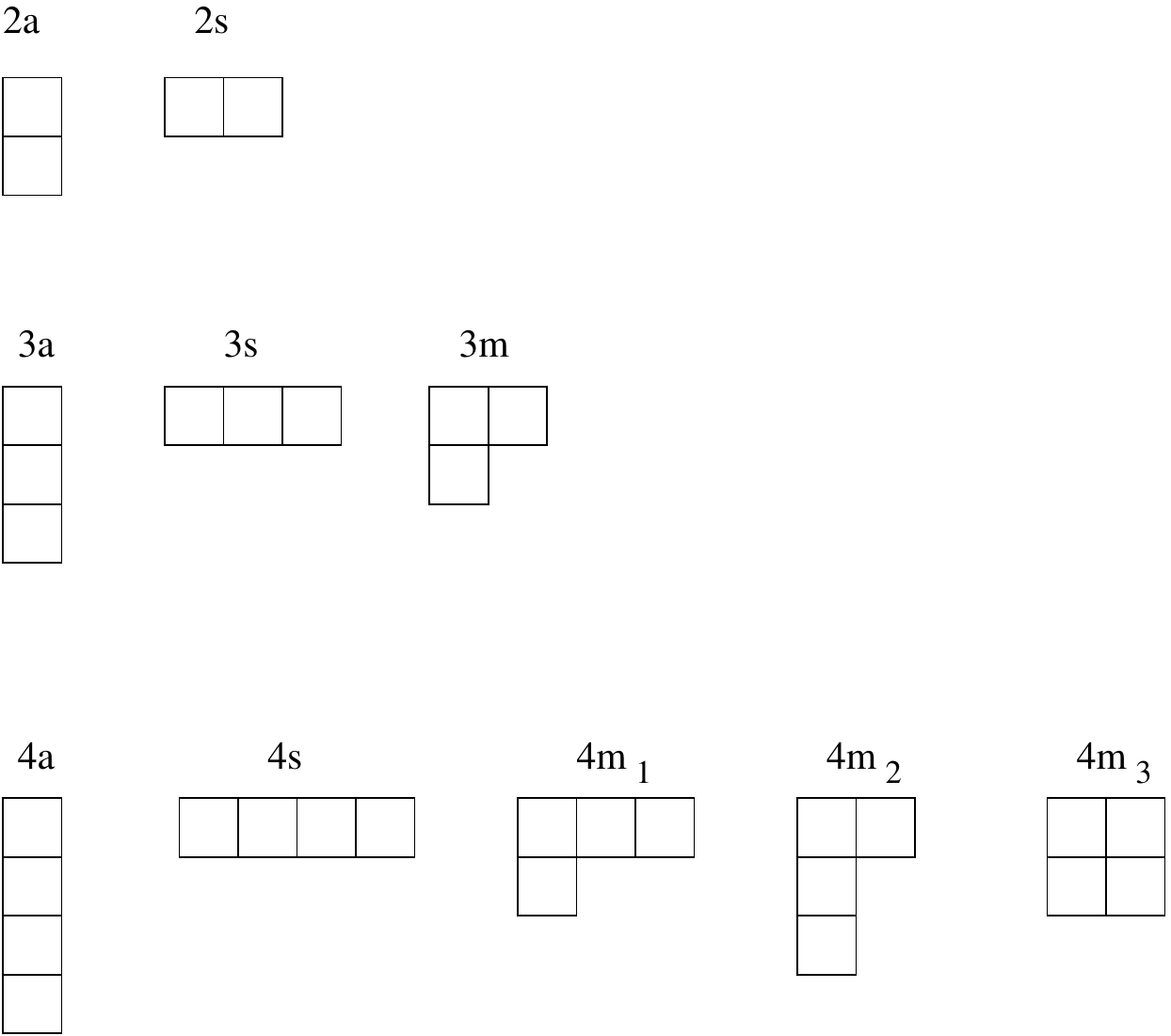}    
\end{center}
\caption{Young tableaux for the decomposition of $(\tr(W))^k$ into irreducible
representations, for $k=2,3,4$.}
\label{young}
\end{figure}

\subsection{k=2}
    The sum runs over the symmetric (2s) and antisymmetric (2a) representations of $N$-ality $k=2$, which are
RC-conjugate.  The decomposition of a product of fundamental representations
into a sum of irreducible representations, as well as the dimensionality and quadratic Casimir of each representation, can be worked out by the usual Young tableau methods (cf.\ e.g.\ Appendices A and C of Ref.\ \cite{Lucini:2001nv}, and Chapter 4 of the text by Cheng and Li \cite{Cheng:1985bj}).  In the especially simple case of $k=2$, we have
\beq
                    \tr_{2a}[U]  +  \tr_{2s}[U]   =  (\tr[U])^2  \ ,
\eeq
with dimensionalities
\beq
            d_{2a} = {N^2 - N \over 2}   ~~~,~~~  d_{2s} = {N^2 + N \over 2} \ ,
\eeq
and ratios of quadratic Casimirs
\beq
           {C_{2a} \over C_F} = {2(N-2) \over N-1} ~~~,~~~  {C_{2s} \over C_F} = {2(N+2) \over N+1} \ .
\eeq
Notice that the dimensionalities and Casimir ratios of the 2a and 2s representations go into one another
under $N\ra -N$.  This is what is required to satisfy the cancellation condition~\eqref{eq:evenodd}.
Then we have
\begin{flalign}
& {1\over N^k} \sum_i g_{R_i^k} d_{R_i^k} \exp[-\s_r A]
\non \\
& = \oh( e^{-\s_{2s}A} + e^{-\s_{2a}A}) + {1 \over 2N} (e^{-\s_{2s}A} - e^{-\s_{2a}A}) \ .
\end{flalign}
Defining
\beq
              \overline{\s} &\equiv& \oh( \s_{2s} + \s_{2a})
\non \\
                                 &=& \oh \left( {2(N+2) \over N+1} + {2(N-2)\over N-1} \right) \s
\non \\
                                 &=& (2 - O(1/N^2)) \s
\eeq
and
\beq
\D \s \equiv  \s_{2s} - \s_{2a} = {4 N \over N^2 - 1} \s \ ,
\eeq
we find 
\begin{flalign}
& {1\over N^k} \sum_i g_{R_i^k} d_{R_i^k} \exp[-\s_r A] 
\non \\
& = e^{-\os A} \left( \cosh(\oh \D\s A) - {1\over N} \sinh(\oh \D\s A) \right)  \ .
\label{k2a}
\end{flalign}
Observe that the series expansion of $\os$ is even in $1/N$, while $\D\s$ contains only odd powers of $1/N$.  From this it follows that the leading strong coupling contributions combine to give
\beq 
\lefteqn{ {1\over N^k} \sum_i g_{R_i^k} d_{R_i^k} \exp[-\s_{R_i^k} A]  }
\non \\
& & = e^{-2\s A} + \mbox{even powers of~} 1/N \ ,
\label{k2}
\eeq
as required by the large-$N$ condition~\eqref{condition}.

\subsection{k=3}  

In this case there is a symmetric (3s) and antisymmetric (3a) representation, which are RC-conjugate, and a mixed (3m) representation which is RC self-conjugate, with Young tableaux
shown in Fig.~\ref{young}.  We have
\beq
 \tr_{3a}[U]  + \tr_{3s}[U]  + 2   \tr_{3m}[U]  &=&  (\tr[U])^3  \ ,
\eeq
with dimensionalities
\beq
d_{3a} &=& {1\over 6} N(N-1)(N-2) \ ,
\non \\
d_{3s} &=& {1\over 6} N(N+1)(N+2) \ ,
\non \\
 d_{3m} &=& {1\over 3} N(N^2-1) \ ,
\eeq
and quadratic Casimirs
\beq
 {C_{3a} \over C_F} = {3(N-3) \over N-1} ~~,~~ {C_{3s} \over C_F} = {3(N+3) \over N+1} \ ,
\non \\
{C_{3m} \over C_F} = {3(N^2-3) \over N^2-1} \ .
\non \\
\eeq
Once again, observe that the dimensionalities (divided by $N^3$) and Casimir ratios of the 3a and 3s representations are interchanged by $N \ra -N$, while these same quantities in the 3m, RC self-conjugate representation are unaffected.   This is what is needed to satisfy~\eqref{eq:evenodd}, and we have, to leading order in the strong coupling expansion
\begin{flalign}
&{1\over N^k} \sum_i g_{R_i^k} d_{R_i^k} \exp[-\s_r A]
\non \\
& = {1\over 6}\left\{ \left(1-{3\over N} +{2\over N^2} \right) e^{-\s_{3a} A} + 
              \left(1+{3\over N} + {2\over N^2} \right)e^{-\s_{3s} A}  \right.
\non \\
& \qquad \left. +  
              4\left(1 - {1\over N^2}\right) e^{-\s_{3m} A} \right\} \ .
\end{flalign}
Define
\beq
\os &\equiv& \oh (\s_{3a} + \s_{3s})
\non \\
&=& \oh \left[ {3(N-3)\over N-1} + {3(N+3)\over N+1} \right] \s
\non \\
&=& \left( 3 - {6\over N^2-1} \right) \s
\non \\
&=& \s_{3m}
\eeq
and
\beq
\D \s &\equiv& \s_{3s} - \s_{3a}
\non \\
&=& \left[{3(N+3)\over N+1} -  {3(N-3)\over N-1}  \right] \s
\non \\
&=& {12 N \over N^2-1}  \s \ .
\eeq
Then
\beq
\lefteqn { {1\over 6}\left\{ \left(1-{3\over N} +{2\over N^2} \right) e^{-(\s_{3m} - \oh \D \s) A} \right. }
\non \\ 
& & \qquad + \left(1+{3\over N} + {2\over N^2} \right) e^{-(\s_{3m} + \oh \D \s) A}
\non \\
& & \qquad + \left. 4\left(1 - {1\over N^2}\right) e^{-\s_{3m} A} \right\}  
\non \\
& & =  {1\over 6}e^{-\s_{3m} A} \left\{ 4\left(1 - {1\over N^2}\right) \right.
\non \\
& & \qquad + 2\left(1 + {2\over N^2}\right)\cosh(\oh \D\s A)
\non \\
& & \qquad \left. - {6\over N} \sinh(\oh \D\s A) \right\}  \ .
\non \\
\eeq
As before, $\os=\s_{3m}$ is even in powers of $1/N$, while $\D \s$ contains only odd powers of $1/N$.  As a consequence, the above expression contains only
even powers of $1/N$, and we have found that the leading strong coupling diagrams give
\beq 
 \lefteqn{ {1\over N^k} \sum_i g_{R_i^k} d_{R_i^k} \exp[-\s_{R_i^k} A] }
\non \\
    & & = e^{-3\s A} + \mbox{even powers of~} 1/N
\eeq
also for $k=3$.
 
\subsection{k=4}
The last example is $k=4$, with RC-conjugate pair 4a and 4s, a second RC-conjugate pair $4m_1$ and $4m_2$, and
a single RC self-conjugate representation $4m_3$.  We have  
\begin{flalign}
& \tr_{4s}U + \tr_{4a}U + 3(\tr_{4m_1}U + \tr_{4m_2}U) + 2\tr_{4m_3}U \nonumber \\
& = (\tr U)^4 \ ,
\end{flalign}
with dimensionalities
\beq
d_{4s} &=& {1\over 24} N(N+1)(N+2)(N+3)  \ ,
\non \\
d_{4a} &=& {1\over 24} N(N-1)(N-2)(N-3) \ ,
\non \\
d_{m_1} &=& {1\over 8} N(N+1)(N-1)(N+2) \ ,
\non \\
d_{m_2} &=& {1\over 8} N(N+1)(N-1)(N-2) \ ,
\non \\
d_{m_3} &=& {1\over 12} N(N+1)(N-1)N \ ,
\eeq
and quadratic Casimirs
\beq
           C_{4s} &=& 2N + 6 - {8\over N} ~~,~~ C_{4a} = 2N - 6 - {8\over N} \ ,
\non \\
           C_{4m_1} &=& 2N + 2 - {8\over N} ~~,~~ C_{4m_2} = 2N - 2 - {8\over N} \ .
\non \\
           C_{4m_3} &=& 2N - {8\over N}
\eeq
Dividing the quadratic Casimirs by the fundamental Casimir $C_F = 2N/(N^2-1)$, and the dimensionalities by $N^4$, the conditions~\eqref{eq:evenodd} can be checked by inspection.  Now define
\beq
\os &\equiv& \oh (\s_{4a} + \s_{4s})
\non \\
&=& \oh (\s_{4m_1} + \s_{4m_2})
\non \\
&=& \s_{4m_3} 
\non \\
 &=& \left( 2N - {8\over N} \right) {2N \over N^2-1} \s
\non \\
&=& \left(4 + \mbox{even powers of }{1\over N}\right) \s
\eeq
and
\beq
\D \s_{as} &\equiv& \s_{4s} - \s_{4a} = {24N \over N^2 -1} \s \ ,
\non \\
\D \s_{12} &\equiv& \s_{4m_1} - \s_{4m_2} = {8N \over N^2 -1} \s  \ .
\eeq
Then we have
\begin{widetext}
\beq
\lefteqn{ {1\over N^k} \sum_i g_{R_i^k} d_{R_i^k} \exp[-\s_r A] }
\non \\
& & \qquad  ={1\over N^4}\left\{ d_{4s} e^{-\s_{4s} A} + d_{4a} e^{-\s_{4a} A} + 3\Bigl(d_{4m_1} e^{-\s_{4m_1} A} + 
      d_{4m_2}e^{-\s_{4m_2} A}\Bigr) + 2d_{4m_3} e^{-\s_{4m_3} A} \right\}
\non \\
& & \qquad  =  {1\over 24 N^3} \left\{ (N^3 + 6N^2 + 11N + 6) e^{-(\os + \oh \D \s_{as})A} 
                                            +  (N^3 - 6N^2  + 11N - 6) e^{-(\os - \oh\D \s_{as})A} \right\}
\non \\
& & \qquad \qquad \qquad + {3\over 8N^3}\left\{ (N^2-1)(N+2) e^{-(\os + \oh\D\s_{12}) A}
       +  (N^2-1)(N-2) e^{-(\os - \oh\D\s_{12}) A} \right\}
 + {1\over 6N^2} (N^2-1) e^{-\os A} 
\non \\
& & \qquad  = e^{-\os A} \left\{ {1\over 12}\left(1 + {11\over N^2}\right)\cosh(\oh\D \s_{as} A) -
 {1\over 2}\left({1\over N} + {1\over N^3}\right)\sinh(\oh\D \s_{as} A) \right.
\non \\
& & \qquad \qquad \qquad \left. + {3\over 4}\left(1-{1\over N^2}\right)\cosh(\oh\D \s_{12} A) 
- {3\over 2}\left({1\over N}-{1\over N^3}\right)\sinh(\oh\D \s_{12} A)
+ {1\over 6}\left(1-{1\over N^2}\right) \right\} \ .
\eeq
\end{widetext}

    Once again, we observe that $\os$ involves only even powers of $1/N$, with $\os=4\s$ at zeroth order, and that the expansion of $\D \s_{as}, \D \s_{12}$
contains only odd powers of $1/N$.  From this we conclude that
\beq 
 \lefteqn{ {1\over N^k} \sum_i g_{R_i^k} d_{R_i^k}\exp[-\s_{R_i^k} A]  }
\non \\
& & = e^{-4\s A} + \mbox{even powers of~} 1/N
\eeq
from the leading strong coupling diagrams, also in the $k=4$ case.

\subsection{Pairwise cancellation at any $k$}

    The examples given above are all illustrations of the cancellation of $1/N$ factors between pairs of irreducible representations, whose Young tableaux
are RC-conjugate pairs.  Let $\s_R^\pm$ denote the part of the $1/N$ expansion of 
$\s_R$ containing even (+) and odd (-) powers of $1/N$ respectively.  Likewise, let 
\beq
            \left[ {g_{R_i^k} d_{R_i^k} \over N^k} \right]^\pm
\eeq
refer to the pieces of the $1/N$ expansion of the quantity in brackets containing even/odd
powers of $1/N$.  Then we see that for the RC-conjugate pairs $(2a,2s)$,
$(3a,3s)$, $(4a,4s)$, and $(4m_1,4m_2)$,
\beq
\left[ {g_R d_R \over N^k} \right]^\pm &=& \pm \left[ {g_{\widetilde{R}} d_{\widetilde{R}} \over N^k} \right]^\pm \,
\non \\
\s_R^\pm &=& \pm \s_{\widetilde{R}}^\pm  \ ,
\label{conjcancel}
\eeq
while for the RC-selfconjugate diagrams $3m$ and $4m_3$, 
\beq
          \s_R^- = \left[ {g_R d_R \over N^k} \right]^- = 0 \ ,
\label{selfconj2}
\eeq
These conditions are equivalent, at leading order in the strong coupling expansion, to~\eqref{eq:evenodd}, and guarantee cancellation of odd powers of $1/N$.

   In Appendix~\ref{appendix:characters} it is shown, for RC-conjugate representations $R$ and $\widetilde{R}$,
that the combination
\beq
      {1\over N^k} \Bigl( \langle \tr_{R}U \rangle + \langle \tr_{\widetilde{R}}U \rangle \Bigr)
\eeq
has an expansion in only even powers of $1/N$.  Hence, given that multiplicities $g_R=g_{\widetilde{R}}$, whatever rule is proposed for $\langle \tr_R(U) \rangle$ must be compatible with the pairwise cancellation of odd powers of $1/N$ 
between  RC-conjugate pairs.  We have seen that Casimir scaling satisfies this rule for $k=2,3,4$, via the relationships~\eqref{conjcancel} among the RC-conjugate pairs.  In Appendix~\ref{appendix:casimirs} we provide a general
proof that the Casimir scaling rule satisfies Eq.~\eqref{conjcancel} for \emph{all} RC-conjugate pairs $(R,\widetilde{R})$ of a given $N$-ality $k$, while Eq.~\eqref{selfconj2} holds for all RC-selfconjugate representations.  Specifically, it is shown in Appendix~\ref{appendix:casimirs} that, in the notation of Eq.~\eqref{pm}, the quadratic Casimirs of RC-conjugate pairs are related by
\beq
\label{eq:conjcasimirs_0}
(C_R/C_F)^{\pm} = \pm (C_{\widetilde{R}} /C_F)^\pm\ ,
\eeq
which means that the Casimir-scaling energies of static $q\overline{q}$ sources separated by a distance $r$ have the same property, i.e.
\beq
\label{eq:conjcasimirs}
E_{r R}^\pm = \pm E_{r \widetilde{R}}^\pm \ .
\eeq
It can also be shown that the ratio of the dimensions over $N^k$ satisfies
\beq
\label{eq:conjdimensions}
\left(\frac{d_{R}}{N^k}\right)^\pm = \pm \left(\frac{d_{\widetilde{R}}}{N^k}\right)^\pm \ .
\eeq
while the multiplicities $g_R=g_{\widetilde{R}}$ of RC-conjugate representations are equal.  For RC-selfconjugate representations, $C_R^- = (d_R/N^k)^- = 0$.  Hence the odd parts in $N$ in the amplitudes come with opposite
signs. Together with Eq.~\eqref{eq:conjcasimirs}, this implies that the odd powers in
$1/N$ cancel between representations corresponding to RC-conjugate
tableaux, while for RC-selfconjugate representations the expansion contains only even powers of $1/N$.     

     Before proceeding, we recall that at leading order in the strong-coupling expansion of the heat-kernel
action, Wilson loops  have the group representation dependence  $\langle \tr_R U \rangle = d_R \exp[-(C_R/C_F) \s]$ (see Eqs.\ (\ref{leading},\ref{cas})), i.e.\ Casimir scaling and a prefactor equal to the representation dimension,
as in two-dimensional continuum gauge theory.  The cancellation of odd powers of $1/N$ among RC-conjugate representations also applies, of course, to any model of gauge theory at large distances which
shares these properties, such as the dimensional reduction conjecture
\cite{Greensite:2003bk},  and the stochastic vacuum theory
\cite{DiGiacomo:2000va}.   In the latter theory, the product of two
Wilson loops in the fundamental representation has been calculated by
Shoshi et al.\ \cite{Shoshi:2002rd}, and in the case of coinciding
loops their expression can be rewritten precisely in the form of Eq.\
\eqref{k2a} above. This mechanism for the cancellation of the odd
powers of $1/N$ is independent of screening, which is not present at
this order in the stochastic vacuum model and does not arise in the two-dimensional continuum gauge theory.

\subsection{Higher Orders}

    So far we have only considered the lowest order of the strong coupling expansion contributing to the VEV of a
rectangular Wilson loop in an irreducible representation $r$.  Cancellation of  odd powers in $1/N$ in higher-order contributions to the product loop is to be expected, given the facts that (i) the strong coupling diagrams of lattice gauge theory can be organized in a $1/N^2$ expansion, and therefore the tensor product loop (suitably divided by a factor of $N^k$) has
such an expansion;  while (ii) the product loop can also be expressed as a sum of loops in irreducible representations, each of which (as we have seen) has an expansion in both even and odd powers of $1/N$.  Consistency requires
cancellation of the odd powers in the sum.  It is instructive to check this cancellation is some special cases
which illustrate processes, such as color screening and ``anti-screening'', which are expected to survive in the continuum limit.

    The leading strong coupling diagram contributing to the VEV of a Wilson loop can be thought of as the creation,
propagation, and annihilation of a string in representation $r$ joining the static sources.  At higher orders, however, there are contributions in which the string joining the sources may be in a different representation from the sources themselves, although of the same $N$-ality.     For example, there are diagrams in which a $2a$ string runs between 
$2s$ sources, and vice-versa.   The geometry of the relevant strong coupling diagram is shown in Fig.~\ref{screen}.  The heavy line represents the rectangular loop $C$, of length $L$ and width $W$, and we are interested in computing 
$\langle \chi_{2x}[U(C)]\rangle$, where $2x=2a$ or $2b$.  Around this loop is a tube of plaquettes, all in the fundamental representation.  Inside the plane of the loop there is a rectangular area of length $L-2$ and width $W-2$ filled with plaquettes in representation $2y$,  where $2y=2a$ or $2b$.  The interior area represents propagation of a string in
the $2y$ representation.  

\begin{figure}[ht]
\centering
\subfigure[~ Tube of plaquettes along the  perimeter of the loop.]{
\resizebox{60mm}{!}{\includegraphics{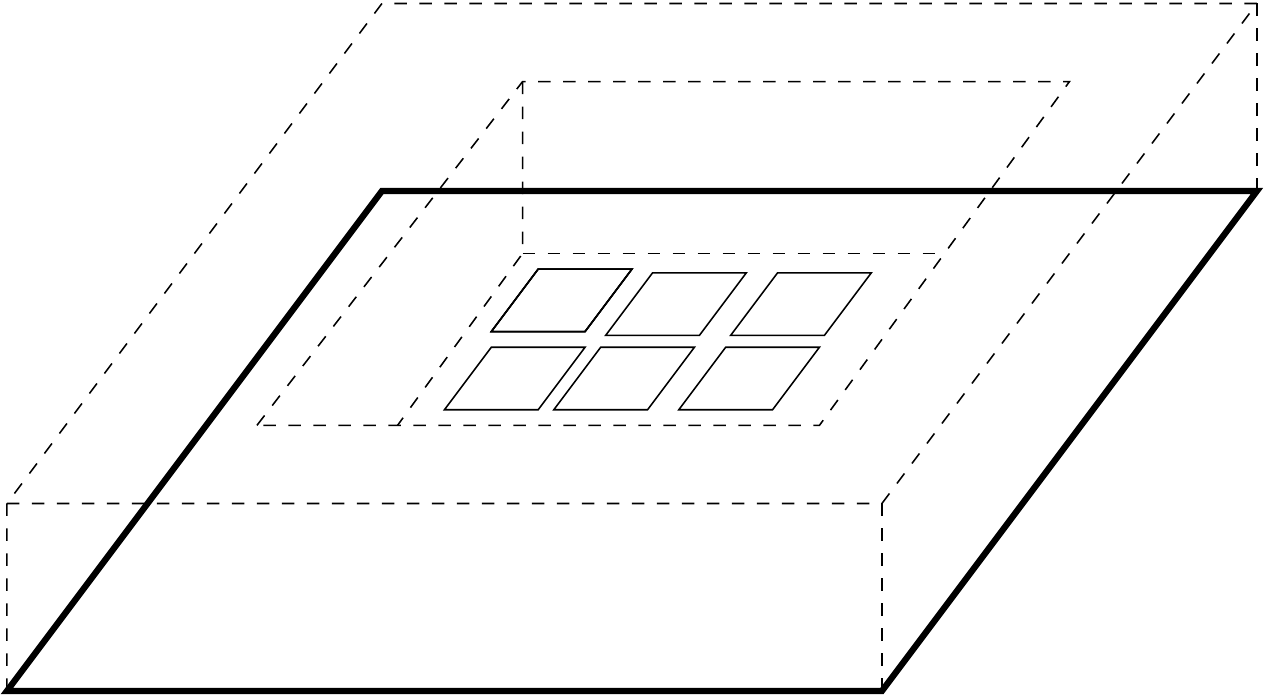}}
\label{screen}
}
\subfigure[~ Details of the tube, outside the plane of the loop.]{
\resizebox{60mm}{!}{\includegraphics{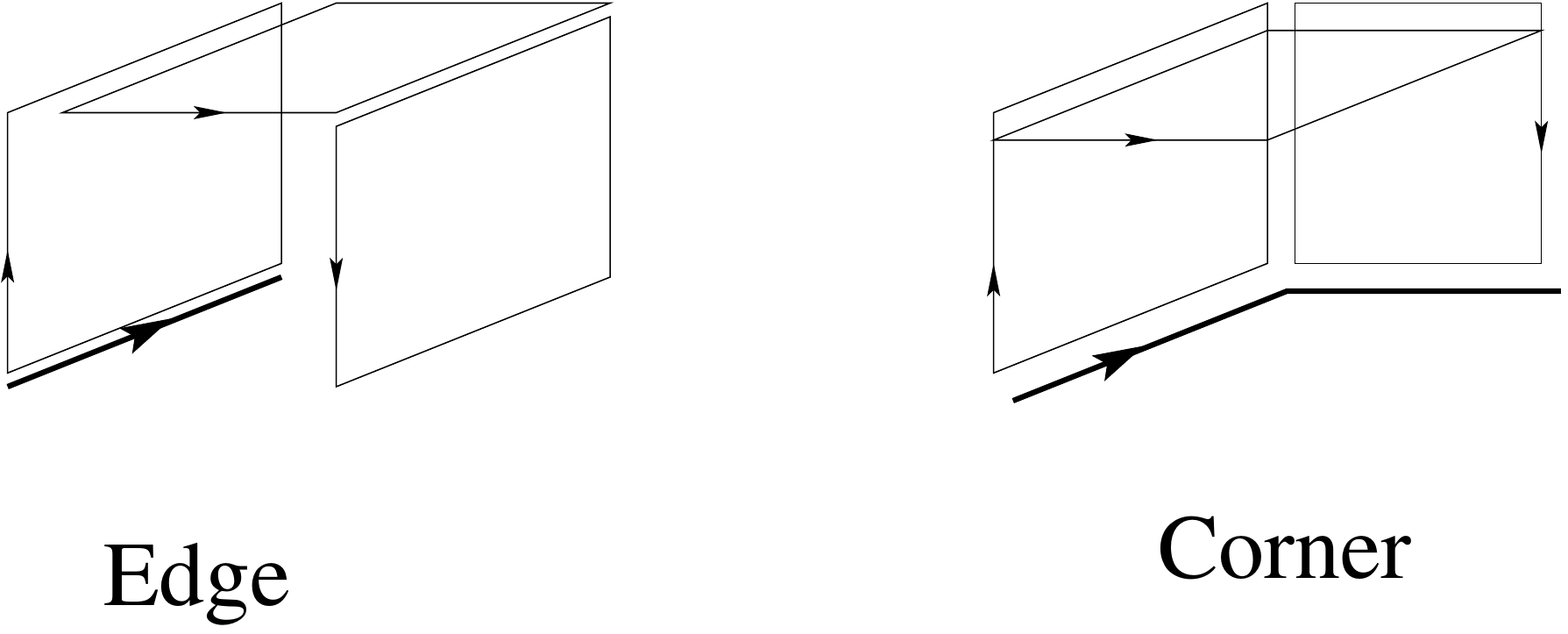}}
\label{sides}
}
\subfigure[~ Fundamental representation plaquettes on the perimeter,  
$2a$ or $2b$ plaquettes in the interior of the loop.]{
\resizebox{60mm}{!}{\includegraphics{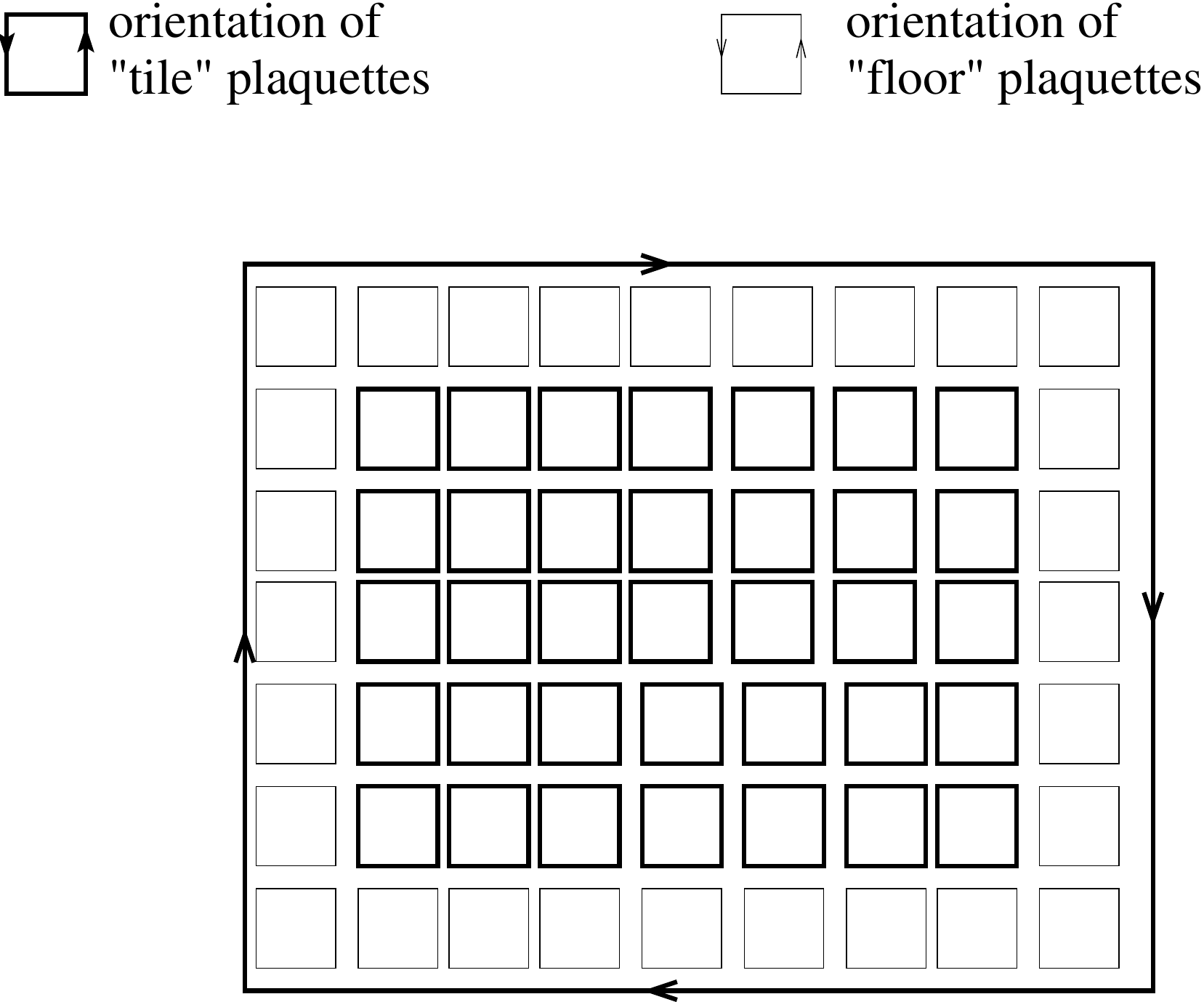}}
\label{tile}
}
\label{fig:subfigureExample}
\caption{Higher-order strong coupling diagram, in which plaquettes in the interior
may be in a different representation than that of the Wilson loop, although of the same $N$-ality.}
\end{figure}

   It is convenient to distinguish between tube plaquettes which are outside the plane of loop $C$, and those which are inside.  The ``outside'' plaquettes of the tube are shown in Fig.~\ref{sides}, and the configuration depends on whether
the tube plaquettes are at a corner or not.   The inside (or ``floor'') plaquettes of the tube run around the inner perimeter of loop $C$, and the remaining minimal area is tiled with plaquettes in representation $2y$, as shown in Fig.~\ref{tile}.

Straightforward integration over link variables then leads to the result
\beq
    \langle \chi_{2a}[U(C)] \rangle &=&  d_{2a} e^{-\s_{2a} A} +   4(D-2) e^{-4\s (P-4)} 
\non \\
   & & \times \Bigl\{
                d_{2a} e^{-\s_{2a}(A-P+4)} + d_{2s} e^{-\s_{2s}(A-P+4)} \Bigr\}
\non \\
    \langle \chi_{2s}[U(C)] \rangle &=&  d_{2s} e^{-\s_{2s} A} + 4(D-2) e^{-4\s (P-4)} 
\non \\
   & & \times \Bigl\{
                d_{2a} e^{-\s_{2a}(A-P+4)} + d_{2s} e^{-\s_{2s}(A-P+4)} \Bigr\}
\non \\ 
\label{result}               
\eeq
where $A$ is the loop area, $P$ is the loop perimeter, and $D$ is the dimension of spacetime.\footnote{The factor
$4(D-2)$ is due to the fact that the tube sticks out in a direction orthogonal to the minimal surface, and there are
$D-2$ possible orthogonal directions.  Then the tube may be above or below the plane of the loop (factor of two), and in either case the floor plaquettes may be in the plane of the loop, or else displaced by one lattice spacing in the orthogonal direction, for an additional factor of two.}  In each case we have summed over the possible representations
$2y=2a,2s$ for the interior plaquettes, and in this sum the odd powers of $1/N$ cancel.

   It is worth noting that the correction term, while of higher order in strong coupling
due to the ``tube'' contribution, is \emph{not} suppressed, relative to the leading term, by any additional powers of
$1/N^2$.  This is in contrast to the strong coupling result for a
Wilson loop in the adjoint representation, where a ``tube'' diagram
which represents string-breaking, and is described by a perimeter law,
is suppressed by a power of $1/N^2$ relative to the area-law
contribution.  There is no such suppression in the conversion between
$2a$ and $2s$ strings, whose representations have the same dimension
in the large-$N$ limit (in contrast to the adjoint and singlet
representations).\footnote{Actually, the tube contribution comes with
an overall factor of the dimension of the interior tiling
plaquettes, and in this sense the tube diagram for color screening
of the adjoint representation, in which the interior can be
considered to be plaquettes in the singlet representation, follows
the same rule as the $N$-ality=2 example we have shown here.}  

\section{$k$-strings in the closed string channel} 
\label{sect:closed}
Wilson loops, in irreducible group representations, probe the
spectrum of open strings with static quarks and antiquarks, in
definite representations $R$ and $\overline{R}$, at the 
ends of the string.  It is not surprising that the spectrum of such
open strings depends on the  representation. However, in the
infinite-$N$ limit the spectrum only  depends on the $N$-ality.  To be
specific, consider the strings with static quarks in the $2s$ and $2a$
representations.  Wilson loops, in these two representations, are
indistinguishable from the product loop in the $N=\infty$ limit (a
consequence of large-$N$ factorization), and hence the spectra of the
two types of open strings must be degenerate in that limit. 
The situation is perhaps a little different for closed strings, which
cannot be characterized by the representation of
quarks at the endpoints. Hence, in this case one would expect a
spectrum which depends only on the $N$-ality also at finite $N$.

    States in the closed string sector contribute to correlators of
Polyakov loops winding around a finite volume in a spacelike
direction.  Let us consider the correlator of two Polyakov loops
in an arbitrary representation $R$, winding around a spatial
compact dimension (for instance $z$) and separated by a temporal
distance $\tau$: 
\begin{gather}
\langle P_R^\dag(0) \sum_{x,y} P_R(x,y,\tau) \rangle = \sum_{n} |\beta_n(r)|^2 e^{-E_n(r)\tau} \ .
\end{gather}
At the lowest order in strong coupling, the only contributing diagram is a sheet of plaquettes in the representation $R$ connecting the Polyakov loops in $(0,0,0)$ and $(0,0,\tau)$:
\begin{gather}
\langle P_R^\dag(0) \sum_{x,y} P_R(x,y,\tau) \rangle = e^{-\sigma_R r
  \tau} \ .
\end{gather}
In the $k=2$ sector, this approximation gives rise to two closed string states $|2a\rangle$ and $|2s\rangle$ defined as:
\begin{gather}
|R\rangle = \frac{1}{L} \sum_{x,y} P_{R}(x,y,\tau=0) |0\rangle \ ,
\end{gather}
($L$ is the common linear size of the $x$ and $y$ directions and takes care of the normalization) with energies respectively $E_{2a}(r) = \sigma_{2a}r$ and $E_{2s}(r) = \sigma_{2s}r$. We want to point out two problems with this lowest level of approximation.

First of all, as for the open string states, color screening is absent in these lowest-order diagrams of the strong coupling expansion. In the closed string channel, color screening is related to the fact that $N$-ality is expected to be the only relevant quantum number. In the absence of an extra selection rule, all the closed string states with charge $2$ under center symmetry should propagate in both the $2s$ and $2a$ channels. But in the lowest order strong coupling diagrams, closed string energy eigenstates, like their open string counterparts, appear to be representation (and not just $N$-ality) dependent. 

The second problem is more subtle. The two closed string states become degenerate at infinite $N$. On the one hand this fact may seem desirable, since, when the correlators of Polyakov loops in the $2a$/$2s$ representations are summed up, the odd powers of $1/N$ must cancel out.  This cancellation follows from the identity
\beq
\lefteqn{\langle P_{2s}^\dag(0) P_{2s}(\tau) \rangle + \langle P_{2a}^\dag(0) P_{2a}(\tau) \rangle =}
\non \\
& &\qquad  \frac{1}{2} \langle P_F^{2\dag}(0) P_F^2(\tau) \rangle + \frac{1}{2} \langle Q_F^\dag(0) Q_F(\tau) \rangle \ ,
\label{identity}
\eeq
where $Q_F$ is the Polyakov line winding twice around the compact direction. One could argue that the odd-power cancellation will happen here in the same way as for the Wilson loops: in pairs of states that become degenerate at infinite $N$ but are split at finite $N$.  On the other hand, accepting the idea that the closed string spectrum depends only on the $N$-ality, and in the absence of a new global symmetry that emerges at $N=\infty$, this degeneracy at infinite $N$ would seem to be accidental.

We want to show that both these problems are solved by considering higher orders in the strong coupling expansion. The two problems are connected: at the same order in strong coupling, the degeneracy at infinite $N$ and the segregation of closed string states by representation are removed.

The idea is to compute matrix elements of the transfer matrix ($T=e^{-H}$ in lattice units) via the strong coupling expansion, in the subspace spanned by the states $|2a\rangle$ and $|2s\rangle$,  and then diagonalize the matrix in this subspace to extract the low-lying energy eigenstates and eigenvalues.   For $N$-ality $k=2$, one has to consider the following $2 \times 2$ matrix:
\begin{gather}
W_{R,R'} = \langle R | e^{-H} | R' \rangle = \langle P_R^\dag(0) \sum_{x,y} P_{R'}(x,y,\tau=1) \rangle \ .
\label{eq:closed_w}
\end{gather}

We want to understand which entries of this matrix contain odd powers of $1/N$. The off-diagonal entries of the matrix $W$ are (dropping the dependence on $x$ and $y$ for convenience):
\beq
W_{2s,2a}  &=& \frac{1}{4} \sum_{xy} \left\{ \langle P_F^{2\dag}(0) P_F^2(1) \rangle - \langle Q_F^\dag(0) Q_F(1) \rangle \right\} 
\non \\
&=& W_{2a,2s} 
\non \\
&\equiv& w_M \ ,
\label{eq:w_offdiag}
\eeq
and, as we will discuss in a moment, they contain only even powers of $1/N$.
In fact the first term in the sum $\langle P_F^{2\dag}(0) P_F^2(1) \rangle$ is a product of four spacelike loops. Naively, at the leading order in $1/N$, this expectation value would completely factorize, giving a $N^4$ contribution. However, in the confined phase the expectation value of a single Polyakov loop vanishes, and the leading contribution is given by
\begin{gather}
\langle P_F^{2\dag}(0) P_F^2(1) \rangle \simeq 2 \langle P_F^{\dag}(0) P_F(1) \rangle_c^2 \ ,
\end{gather}
which is order $N^0$. We recall that the general leading behaviour for connected expectation values of products of Wilson loops is $\langle W_1 \cdots W_n\rangle_c = O(N^{2-n})$. Since this term starts from $N^0$, and subleading corrections are generally suppressed by powers of $1/N^2$, we conclude that the term $\langle P_F^{2\dag}(0) P_F^2(1) \rangle$ contains only even powers of $1/N$. The same conclusion is found easily to hold also for the second term $ \langle Q_F^\dag(0) Q_F(1) \rangle $ in the sum in Eq.~\eqref{eq:w_offdiag}.

Consider now the diagonal entries of the matrix $W$:
\begin{widetext}
\begin{gather}
W_{2s,2s} = \frac{1}{4} \sum_{xy} \left\{ \langle P_F^{2\dag}(0) P_F^2(1) \rangle + \langle Q_F^\dag(0) Q_F(1) \rangle + 2 \langle P_F^{2\dag}(0) Q_F(1) \rangle \right\} \ , \\
W_{2a,2a} = \frac{1}{4} \sum_{xy} \left\{ \langle P_F^{2\dag}(0) P_F^2(1) \rangle + \langle Q_F^\dag(0) Q_F(1) \rangle - 2 \langle P_F^{2\dag}(0) Q_F(1) \rangle \right\} \ ,
\end{gather}
\end{widetext}
While the first two terms in the sum contain only even powers of $1/N$ and appear with the same coefficients in the two entries, we will see that the third term contains only odd powers of $1/N$. In fact $\langle P_F^{2\dag}(0) Q_F(1) \rangle$ would naively be order $N^3$, but because of the center symmetry its leading order coincides with the connected expectation value which is $1/N$. Again, since subleading corrections are generally suppressed by powers of $1/N^2$, we conclude that the term $\langle P_F^{2\dag}(0) Q_F(1) \rangle$ contains only odd powers of $1/N$. Moreover this term appears with a different sign in the diagonal entries, which are hence related to each other via the $N \to -N$ transformation. We introduce the notation:
\begin{gather}
W_{2s,2s} = w_1 - \frac{w_2}{N} \ , \\
W_{2a,2a} = w_1 + \frac{w_2}{N} \ ,
\end{gather}
where both $w_1$ and $w_2$ contain only even powers of $1/N$.

\begin{figure}[t!]
\centering 
\includegraphics[width=.5\textwidth]{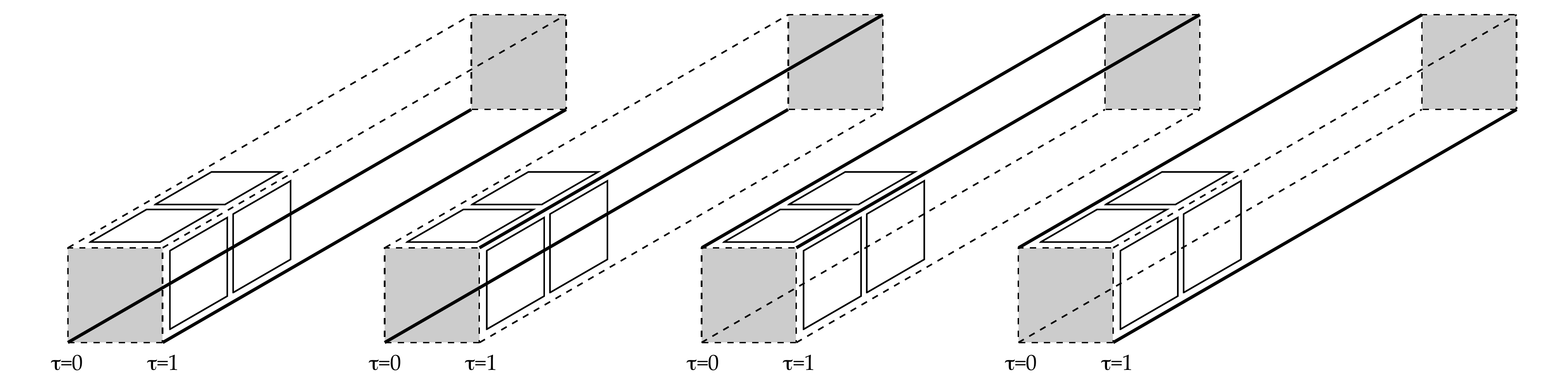}
\caption{Diagrams contributing to $w_M$. The tubes run through the periodic lattice in the $z$-direction, and the two gray areas in each tube are identified.  The thicker lines represent the two Polyakov loops, one in the $2s$ and the other in the $2a$ representation. The two Polyakov loops have opposite orientations. The tube is tiled with plaquettes in the (anti)fundamental representations.}
\label{fig:poly1}
\end{figure}

\begin{figure}
\centering
\includegraphics[width=.3\textwidth]{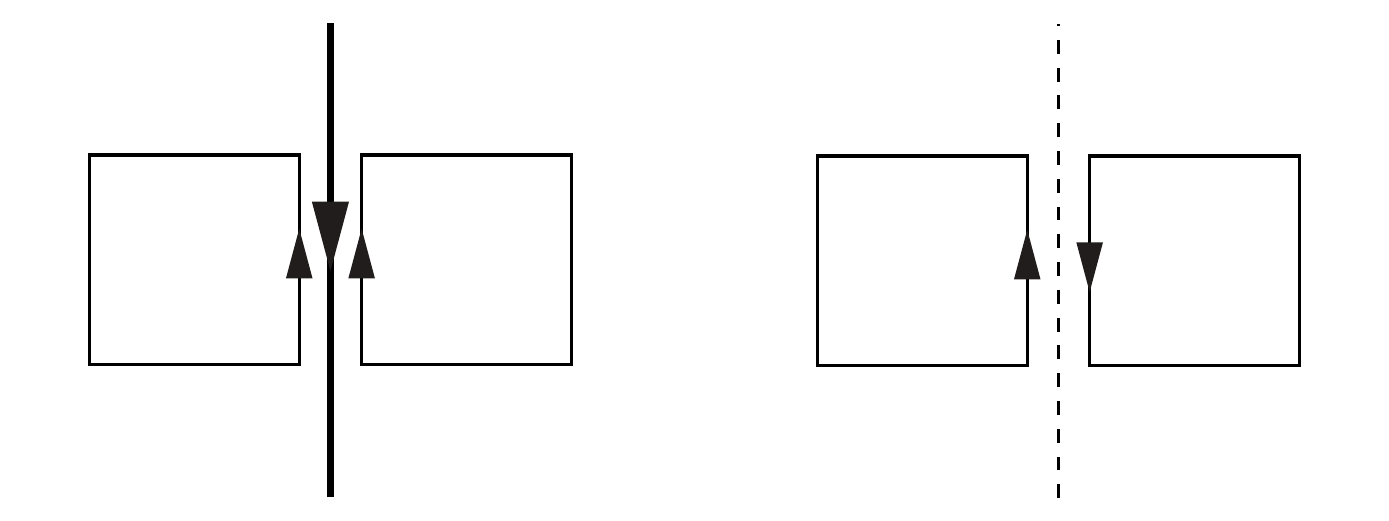}
\caption{Detail showing the orientation of Polyakov loops and plaquettes in the diagrams contributing to $w_M$ (see Fig.~\ref{fig:poly1}).}
\label{fig:poly2}
\end{figure}

The splitting at infinite $N$ happens when the off-diagonal entry $w_M$ is non vanishing. The lowest-order contribution is given by a tube of plaquettes in the (anti)fundamental representation connecting the two Polyakov loops (Figs.~\ref{fig:poly1} and~\ref{fig:poly2}):
\begin{gather}
w_M(r) = 12 e^{-4\sigma r} \ .
\label{eq:wm}
\end{gather}
The energy eigenstates $|L\rangle$ and $|H\rangle$ can be found by diagonalizing the matrix $2\times 2$ submatrix
$W$ of the transfer matrix $T$:
\begin{gather}
|H\rangle = \cos \omega |2s\rangle - \sin \omega |2a\rangle \label{eq:eigH} \ ,\\
|L\rangle = \sin \omega |2s\rangle + \cos \omega |2a\rangle \label{eq:eigL} \ , \\
\cos^2 \omega = \frac{w_M^2}{\left[ \sqrt{\frac{w_2^2}{N^2} + w_M^2} - \frac{w_2}{N} \right]^2 + w_M^2} \label{eq:cos2} \ , \\
\sin^2 \omega = \frac{w_M^2}{\left[ \sqrt{\frac{w_2^2}{N^2} + w_M^2} + \frac{w_2}{N} \right]^2 + w_M^2} \label{eq:sin2} \ .
\end{gather}
The eigenvalues of $W$ are $e^{-E_L}$ and $e^{-E_H}$, where the exponents are the energies of the perturbed states:
\begin{gather}
E_L(r) = - \log \left\{ w_1 + \sqrt{\frac{w_2^2}{N^2} + w_M^2} \right\} \ , \\
E_H(r) = - \log \left\{ w_1 - \sqrt{\frac{w_2^2}{N^2} + w_M^2} \right\} \ .
\label{ELH}
\end{gather}

These last two formulae contain the central result of this section. A few comments are now in order.

    In the first place, the quantities $w_1$, $w_2$, $w_M$ contain only even powers of $1/N$, and the same is true also for the closed string energies.  One might then confidently predict, from Eq.~\eqref{ELH} that the asymptotic string tension will also contain only even powers of $1/N$.   On the other hand, at fixed $N$ and large compactification radius $r$, the leading contributions to the entries of the matrix $W$ are given by the largest exponential and therefore by the lowest order in strong coupling. At this level of approximation $W_{2s,2s} = \langle P_{2s}^\dag(0) \sum_{x,y} P_{2s}(x,y,\tau=1) \rangle$ is obtained by picking only the term at $x=y=0$ and by tiling the gap between the two Polyakov loops with plaquettes in the $2s$ representation, which gives $W_{2s,2s} \simeq e^{-\sigma_{2s}r}$. In the same way we get $W_{2a,2a} \simeq e^{-\sigma_{2a}r}$.  The off-diagonal term  $w_M$ is negligible in the fixed $N$, large $r$ limit,
so we have
\begin{gather}
w_1(r) \simeq \frac{e^{-\sigma_{2a}r} + e^{-\sigma_{2s}r}}{2} \ , \\
w_2(r) \simeq N \frac{e^{-\sigma_{2a}r} - e^{-\sigma_{2s}r}}{2} \ , \\
w_M(r) \simeq O(e^{-4\sigma r}) \ .
\end{gather}
Inserting these expressions into the energies, we find  
\begin{widetext}
\begin{gather}
E_L(r) \simeq -\log \left\{ \frac{e^{-\sigma_{2a}r} + e^{-\sigma_{2s}r}}{2} + \left| \frac{e^{-\sigma_{2a}r} - e^{-\sigma_{2s}r}}{2} \right| \right\} \simeq \sigma_{2a} r \ , 
\non \\
E_H(r) \simeq -\log \left\{ \frac{e^{-\sigma_{2a}r} + e^{-\sigma_{2s}r}}{2} - \left| \frac{e^{-\sigma_{2a}r} - e^{-\sigma_{2s}r}}{2} \right| \right\} \simeq \sigma_{2s} r \ .
\label{ELH1}
\end{gather}
\end{widetext}
Here we seemingly encounter a paradox.  It was asserted that $E_{L,H}(r)$ are functions of $1/N^2$, so how can they
possibly be proportional to $\s_{2a,2s}$, which, in the lattice strong coupling expansion, are known to contain odd powers of $1/N$?

    The ``cosh argument'' of Sect.~\ref{sect:wl} is again instructive.  As noted there,
 $\log \cosh(x)$ at small $x$ has an expansion in powers of $x^2$ only.
On the other hand, at $x \gg 1$, and $x \ll -1$, we have $\log \cosh(x) \approx |x|$, which is \emph{linear} in $x$ for large $x$ of a fixed sign.  The point to notice is that whether we expand at large $x$ or at small $x$, in either case the expansion is even with respect to $x\ra -x$.  The situation is similar, in some ways, in the case at hand.  If we expand the energies at fixed $r$, taking $N$ very large (and not discarding $w_M$), then we will indeed obtain a power series in 
$1/N^2$.  If, on the other hand, we expand at fixed $N$, taking $r$ very large, then the answer is only required to be even under $N \ra -N$, and can, in particular, depend on $|N|$.  The point is that the large $r$, fixed $N$ limit, and the fixed $r$, large $N$ limit need not look at all alike, but in either case the resulting expression is an even function of $N$.  With this in mind, the  correct answer for $E_{L,H}$ in the large $r$ fixed $N$ limit shown above is to replace $\s_{2a,2s}$ on the right hand side of Eq.~\eqref{ELH1} by the quantities $\s'_{2a,2s}$, obtained from the corresponding unprimed quantities by replacing $N$ by $|N|$.

This resolves the question of how the energies can have a $1/N^2$ expansion, yet the asymptotic string tension may have corrections in odd powers of $1/N$ (which we now understand to be odd powers
of $1/|N|$).\footnote{ An interesting question is whether it is possible, in principle, for the leading corrections to the asymptotic string tensions to involve fractional powers of $1/|N|$.   Although we do not believe that such exotic $N$-dependence occurs in pure SU($N$) gauge theories, neither can we readily prove that fractional powers cannot arise in any model.  The requirement is that when we take the large-$N$ limit prior to the large-loop limit, all fractional powers must cancel.}
 
     Finally, we turn to the question of energy degeneracy in the closed string sector. The quantities 
$w_1$, $w_2$, $w_M$ have a finite large-$N$ limit. Separating the large $r$ behaviour (at finite $N$, as discussed in the previous point) in $w_1 = e^{-\sigma_{2a}r} + \Delta w_1 $, energies and mixing angles at infinite $N$ are:
\begin{eqnarray}
\nonumber
E_L(r) &=& - \lim_{N \to \infty} \log( w_1 + w_M ) \\ 
    &=& 2 \sigma r - \log( 1 + \Delta w_1 + 12 e^{-2\sigma r} ) \ , \\
\nonumber
E_H(r) &=& - \lim_{N \to \infty} \log( w_1 - w_M ) \\
    &=& 2 \sigma r - \log( 1 + \Delta w_1 - 12 e^{-2\sigma r} ) \ , 
\end{eqnarray}
\begin{eqnarray}
\cos^2 \omega = \sin^2 \omega = \frac{1}{2} \ .
\end{eqnarray}
As anticipated, the two levels are split even at $N=\infty$, and the splitting is generated by the off-diagonal element of $W$.

The discussion above solves the problem of the accidental degeneracy at infinite $N$, showing that this degeneracy is removed if one goes at a large enough order in the strong coupling expansion. We now find for the correlators of Polyakov loops,  
\begin{eqnarray}
\lefteqn{\langle P_{2a}^\dag(0) \sum_{x,y} P_{2a}(x,y,\tau) \rangle =}
\non \\ 
 & & \qquad \cos^2 \omega \ e^{-E_L(r)\tau} + \sin^2 \omega \ e^{-E_H(r)\tau} \ , 
\label{eq:corr_2a} \\
\lefteqn{\langle P_{2s}^\dag(0) \sum_{x,y} P_{2s}(x,y,\tau) \rangle =}
\non \\
 & & \qquad \sin^2 \omega \ e^{-E_L(r)\tau} + \cos^2 \omega \ e^{-E_H(r)\tau} \ .
\label{eq:corr_2s}
\end{eqnarray}
One can immediately see that, as a result of this mixing (which is generated along with the infinite-$N$ splitting),  both the eigenstates in the $k=2$ sector propagate in both of the correlators, as expected. The stable string tension in the closed string channel is the same for both the correlators and is equal to $\sigma_{2a}$. 

    In the case of Polyakov loops, from Eq.~\eqref{identity} and remarks immediately following, it is clear that
that the sum of correlators in~\eqref{eq:corr_2a} and~\eqref{eq:corr_2s} must have an expansion in powers of $1/N^2$, which
means that any odd powers in this sum must cancel.  However, the odd-power cancellation mechanism is a little different for Polyakov loop correlators, as compared to Wilson loops.  In the Polyakov loop case,
since the states propagating in the two correlators  are the same and the energies are non-degenerate at $N=\infty$,  the energies must have only even-power corrections, as indeed they do. However the amplitudes $\cos^2 \omega$ and $\sin^2 \omega$ contain odd-power corrections (from the $w_2/N$ term in Eqs.~\eqref{eq:cos2} and~\eqref{eq:sin2}) and are connected by the $N \to -N$ transformation. Therefore, when the two correlators are summed up, the cancellation still happens pairwise between the amplitudes of the same state.

As an aside, we note that in the large
$r$ limit (with $N$ kept fixed), Eqs.~\eqref{eq:cos2}
and~\eqref{eq:sin2} yeld
\begin{gather}
\cos^2 \omega =1 \ , \sin^2 \omega = 0 \ ,
\end{gather} 
i.e. the two states $|AS\rangle$ and $|S\rangle$ become respectively the
Hamiltonian eigenstates $|L\rangle$ and $|H\rangle$. The separation of the
closed string eigenstates according to gauge group representations in the large $r$
limit has been suggested in~\cite{Bringoltz:2008nd,Athenodorou:2008cj}. However, in our case this separation
heavily relies on the particular order at which the strong coupling
expansion has been truncated. For this reason, our calculation does
not allow us to infer that the Hamiltonian eigenstates separate by
representation in the large $r$ limit.

\section{Discussion: More on degeneracy}
\label{sect:discussion}
So far we have shown, in the framework of the lattice strong coupling
expansion, that $O(1/N)$ corrections to $k$-string tensions are not
excluded by large-$N$ counting arguments.  As a counter-argument to,
e.g., Refs.~\cite{Armoni:2003ji,Armoni:2003nz}, this example is
sufficient, but of course it does not prove that $k$-strings actually
have $O(1/N)$ corrections in the continuum theory.   On the other hand
some of our results, specifically Eq.~\eqref{eq:evenodd} which is
crucial to the $1/N$ cancellation mechanism, are valid at any
coupling.  Suppose it were true that all energies in the open string
channel, in the limit of fixed $r$ and large $N$, had an expansion in
powers of $1/N^2$ only.  Then Eq.~\eqref{eq:evenodd} would imply a
degeneracy in energy levels found in RC-conjugate representations at
\emph{finite $N$}, and this degeneracy, if it exists, would appear to
be accidental. Degeneracies among open string states with the same
$N$-ality still occur (and indeed that fact is crucial to the
cancellation mechanism), but strictly at $N=\infty$. 

   The work of Ref.\  \cite{KorthalsAltes:2005ph} anticipates our
own.  These authors have also argued for the presence of $O(1/N)$
corrections to the $k$-string tensions, and pointed out the
non-commutativity of the large-$N$ and  
large-$\tau$ limits.  Our arguments are more general, in particular
as regards the pairwise cancellation of odd powers of $1/N$ among
RC-conjugate pairs.  But we also differ
from~\cite{KorthalsAltes:2005ph} on certain points, in particular 
regarding the degeneracy of closed string energy levels at
$N=\infty$.  Ref.\ \cite{KorthalsAltes:2005ph} argues that a
degeneracy exists also in the closed string sector at infinite $N$,
and they attribute this to the fact that $P_F^2$ and $Q_F$ create
orthogonal states of the same energy.  We differ with 
Ref.\ \cite{KorthalsAltes:2005ph} on this point at strong coupling,
and we will now argue that these states are non-degenerate at weak
couplings as well.  

We recall that $P_F$ and $Q_F$ are Polyakov lines in the fundamental representation winding respectively once and twice around the compact dimension. We define the normalized states:
\begin{gather}
| P_F^2 \rangle = \frac{1}{\sqrt{\langle \left| \sum_{xy} P_F^2(x,y,0) \right|^2 \rangle} } \sum_{xy} P_F^2(x,y,0) | 0 \rangle \ , \\
| Q_F \rangle = \frac{1}{\sqrt{\langle \left| \sum_{xy} Q_F(x,y,0) \right|^2 \rangle} } \sum_{xy} Q_F(x,y,0) | 0 \rangle \ . 
\end{gather}
These two states become orthogonal in the large-$N$ limit, and they also evolve in time in orthogonal subspaces:
\begin{flalign}
& \langle P_F^2 | e^{-H\tau} | Q_F \rangle \nonumber \\
& =
\frac{
\langle 0 | \sum_{xy} P_F^{2\dag}(\tau) \sum_{xy} Q_F(\tau=0) | 0 \rangle
}{
\sqrt{\langle \left| \sum_{xy} Q_F \right|^2 \rangle
\langle \left| \sum_{xy} P_F^2 \right|^2 \rangle}
} = O \left( \frac{1}{N} \right)
\ .
\label{eq:orthogonality}
\end{flalign}
Here we used the fact that both $\langle P_F^{2\dag} P_F^2 \rangle$ and $\langle Q_F^\dag Q_F \rangle$ are of order $N^0$, while $\langle P_F^{2\dag} Q_F \rangle$ is of order $1/N$.

Let us take $N=\infty$ from now on. It is useful to decompose each of the states above in energy eigenstates of the closed string with $k=2$:
\begin{gather}
| P_F^2 \rangle = \sum_n \alpha_n |E_n,P^2_F\rangle \ , \\
| Q_F \rangle = \sum_n \beta_n |E_n,Q_F\rangle  \ ,
\end{gather}
where the dependence of $\alpha$, $\beta$ and $E$ on $r$ is understood.
The convention is that each energy level appears once in each decomposition. The eigenstates are normalized, and the two states $|E_n,P_F^2\rangle$ and $|E_n,Q_F\rangle$ have the same energy $E_n$ but they can be different. From the orthogonality relationship~\eqref{eq:orthogonality}
\begin{gather}
\sum_n \alpha_n^* \beta_n \langle E_n, P_F^2 | E_n, Q_F \rangle e^{-E_n \tau} = 0 \ ,
\end{gather}
which is valid at any $\tau$, we must have
\begin{gather}
\alpha_n^* \beta_n \langle E_n, P_F^2 | E_n, Q_F \rangle = 0\ ,
\end{gather}
for each energy level. Two solutions are possible.
\begin{enumerate}
\item \textit{Degeneracy scenario:} $\langle E_n, P_F^2 | E_n, Q_F \rangle = 0$, which means that the energy level is doubly degenerate. This is the scenario assumed in \cite{KorthalsAltes:2005ph}.  The degeneracy would be broken at finite $N$ by $1/N$ corrections in the energy. However, we have already seen that the strong coupling expansion does not produce $1/N$ corrections in the energies (evaluated at fixed $r$, large-$N$), but only corrections in 
powers of $1/N^2$.
\item \textit{Segregation scenario:} $\alpha_n^* \beta_n = 0$, which means that a state with energy $E_n$ is present either in $|P_F^2\rangle$ or in $|Q_F\rangle$, but not in both. This scenario is realized if no degeneracy exists 
at infinite $N$.
\end{enumerate}

In the strong coupling expansion, one can use Eqs.~\eqref{eq:eigH} and~\eqref{eq:eigL} in order to write the states $|P_F^2\rangle$ and $|Q_F\rangle$ in terms of the Hamiltonian eigenvectors $|L\rangle$ and $|H\rangle$:
\begin{eqnarray}
\nonumber
|P_F^2\rangle &=& |2s\rangle + |2a\rangle \\
&=& ( \cos \omega + \sin \omega ) |L\rangle + ( \cos \omega - \sin \omega ) |H\rangle \ , 
\end{eqnarray}
\begin{eqnarray}
\nonumber
|Q_F\rangle &=& |2s\rangle - |2a\rangle \\
&=& ( \cos \omega + \sin \omega ) |H\rangle + ( \sin \omega - \cos \omega ) |L\rangle \ .
\end{eqnarray}
At infinite $N$, since $\cos \omega$ and $\sin \omega$ become equal, we get that $|P_F^2\rangle=|L\rangle$ and $|Q_F\rangle=|H\rangle$. Therefore the strong coupling expansion supports the segregation scenario, in contrast to the argument of~\cite{KorthalsAltes:2005ph}. Once $|P_F^2\rangle$ and $|Q_F\rangle$ are decomposed in eigenstates of the Hamiltonian, the amplitudes of some of those eigenstates go to zero in the large-$N$ limit in such a way that $|P_F^2\rangle$ and $|Q_F\rangle$ become orthogonal to each other. Vanishing amplitudes signals a selection rule that becomes active at infinite $N$.

We want to argue that such a selection rule, valid at least for the
ground state, exists even in the continuum limit. The argument is
based on volume reduction at infinite $N$~\cite{Kovtun:2007py} and large-$N$ factorization.

Again, we keep $N$ exactly equal to infinity. If we call $E_{0}^F(r)$ the ground state in the sector of fundamental closed strings (we write the size of the compact direction explicitly as a subscript to expectation values):
\begin{gather}
\langle P_{F}^\dag(0) \sum_{x,y} P_{F}(x,y,\tau) \rangle_r = A_0(r) e^{-E_0^F(r) \tau} + \dots \ ,
\end{gather}
then the lowest-energy state propagating in the $P_F^2$ channel has energy $2E_0^F(r)$, thanks to factorization:
\begin{flalign}
&\langle P_{F}^{2\dag}(0) \sum_{x,y} P_{F}^2(x,y,\tau) \rangle_r \nonumber \\
&= 2\langle P_{F}^\dag(0) \sum_{x,y} P_{F}(x,y,\tau) \rangle_r^2 \nonumber \\
&= 2A_0(r)^2 e^{- 2E_0^F(r) \tau} + \dots \ .
\end{flalign}
This formula means that the operator $P_F^2$ creates two non-interacting fundamental closed strings at infinite $N$.

On the other hand the operator $Q_F$ creates a fundamental closed string wrapping twice around the compact direction. The precise formulation of this fact relies on volume reduction~\cite{Kovtun:2007py}. Correlators of $Q_F$ around a compact direction of size $r$ can be unfolded and they become correlators of $P_F$ around a compact direction of size $2r$: 
\begin{flalign}
& \langle Q_{F}^{\dag}(0) \sum_{x,y} Q_{F}(x,y,\tau) \rangle_r \nonumber \\
& = 2 \langle P_{F}^\dag(0) \sum_{x,y} P_{F}(x,y,\tau) \rangle_{2r} \nonumber \\
& = 2 A_0(2r) e^{- E_0^F(2r)\tau} + \dots \ .
\end{flalign}
The lowest-energy state propagating in the $Q_F$ channel has energy $E_0^F(2r)$.

Degeneracy of the ground states in the $Q_F$ and $P_F^2$ channels would require that $E_0^F(2r) = 2E_0^F(r)$
exactly, which is satisfied only by linear functions (assuming regularity).  On the contrary, $E_0^F(r)$ is only linear asymptotically, and terms which violate linearity are expected to be present at any finite $r$. In the strong coupling expansion, there will be corrections which decay exponentially with $r$.  In the continuum, strict linearity is broken more strongly by L\"uscher terms.  The existence of such terms means that the degeneracy at $N=\infty$ is broken, and the segregation scenario is favored.

\section{Conclusions}
\label{sect:conclusions}
In this article we have shown that large-$N$ considerations do not necessarily rule out
$k$-string tensions whose $1/N$ expansions contains odd, as well as even, powers of the
expansion parameter.  The large-$N$ expansion does require that certain observables,
such as Wilson loops in the tensor product representation, have an expansion in powers
of $1/N^2$ only, but we have seen that this can be achieved by cancellation
of odd powers of $1/N$, among open string states whose energies are degenerate in the large-$N$ limit.
We have found a concrete example in which these conjectured $1/N$ cancellations really do occur, 
namely, strong coupling lattice gauge theory with the heat-kernel
action.  

In the same example we have seen that closed string states propagating in spacelike Polyakov loop correlators have energies which, in the limit of fixed lattice length $r$ but very large $N$, have an
expansion purely in powers of $1/N^2$.  On the other hand, the limit relevant to the asymptotic string tension is
fixed $N$ but very large $r$, and in this limit we find asymptotic string tensions with $1/N$ (or more precisely,
$1/|N|$) corrections.  As we have discussed in detail, some our results are general enough extend beyond the strong coupling computation presented in this article.  It is important to stress again that the large-$N$ and large-loop limits do not commute, and an expansion in powers $1/N^2$ requires that the former limit is taken first.

An implication of our work is that Casimir scaling, which is realized in
the strong coupling model, is in principle compatible with the
large $N$ expansion. Nothing in our present article suggests that Casimir scaling
is actually preferred over, e.g., the sine law in the continuum theory.
That is a dynamical issue.   We have only argued
that Casimir scaling, or a similar behavior with a $1/N$ (as opposed to $1/N^2$) expansion for the
$k$-string tensions, cannot be automatically ruled out on the grounds of the large-$N$ expansion.

It is important to know whether the leading corrections to $k$-string tensions really
do begin at $O(1/N)$, or alternatively at $O(1/N^2)$, because the answer should give 
us an important clue about the dynamics of confinement.  At present, the lattice Monte Carlo 
data favors $O(1/N)$ \cite{Bringoltz:2008nd}, at least for $k=2$ strings in $D=3$ dimensions. 
We believe that the large-$N$ dependence of $k$-string tensions is a subject that
deserves further lattice investigation, perhaps with the help of advanced error-reduction
algorithms \cite{Luscher:2001up}, and the results may serve as a useful constraint
in the further development of theoretical ideas about the confinement mechanism.

\acknowledgments{
This work was begun at the ECT* in Trento, Italy,  during the workshop
``Confining Flux Tubes and Strings''. J.G. and B.L. are indebted to Ofer Aharony,
Adi Armoni, Misha Shifman and the other workshop
participants for enlighting conversations related to the subject of this paper.  
J.G.\ is also grateful
to Barak Bringoltz, for some earlier stimulating discussions on these topics. 
B.L. thanks Tim Hollowood for discussions on properties of Casimir
operators for representations corresponding to RC-conjugate tableaux. B.L. and A.P. thank Adi
Armoni for countless stimulating conversations on $k$-strings at large-$N$ over the last five years. A.P.\ thanks Pat Kalyniak for correspondence on the
symmetric group. We thank A. Armoni, R. Auzzi, B. Bringoltz, M. Kneipp
and H. Neuberger for comments on the manuscript.
The work of B.L. is supported by the Royal Society through the University 
Research Fellowship scheme and by STFC under contract ST/G000506/1.
J.G.'s research is supported in part by 
the U.S.\ Department of Energy under Grant No.\ DE-FG03-92ER40711.
This research was supported in part by the European Community - Research Infrastructure Action
under the FP7 ``Capacities'' Specific Programme, project ``HadronPhysics2''.
}

\appendix
\section{Characters in RC-conjugate representations}
\label{appendix:characters}
Consider the $N$-dimensional vectorial space $V=\mathbb{C}^N$, and the space $V^{\otimes k}$ of the $k$-rank tensors. The group SU($N$) acts on the tensorial space with the reducible representation $R=\Box \otimes \dots \otimes \Box$:
\begin{equation}
U \in \textrm{SU}(N) \quad : \quad R(U) \cdot v_1 \otimes \dots \otimes v_k = U v_1 \otimes \dots \otimes U v_k \ ,
\end{equation}
where $R(U)$ denotes the group element $U$ in the product representation $R$.
The symmetric group $\Sigma_k$ (group of the permutations of $k$ elements) acts also on the tensorial space with the representation $r$:
\begin{equation}
\sigma \in \Sigma_k \quad : \quad r(\sigma) \cdot v_1 \otimes \dots \otimes v_k = v_{\sigma^{-1}(1)} \otimes \dots \otimes v_{\sigma^{-1}(k)} \ .
\end{equation}
Since $[R(U),r(\sigma)]=0$ for each choice of $U$ and $\sigma$, the tensorial space can be decomposed in simultaneous representations of the two groups. Those representations are labelled by the Young tableaux $S$ with $k$ boxes:
\begin{equation}
V^{\otimes k} = \bigoplus_S V_S \ .
\end{equation}
The space $V_S$ is the basis for an irreducible representation $R_S \times r_S$ of the group $\textrm{SU}(N) \times \Sigma_n$, which is given by the product of an irreducible representation $R_S$ of SU($N$) and an irreducible representation $r_S$ of $\Sigma_n$. The representation $R_S \times r_S$ occurs exactly once, which means that the representation $R_S$ of the group SU($N$) occurs with multiplicity $\dim r_S$, and the representation $r_S$ of the group $\Sigma_n$ occurs with multiplicity $\dim R_S$.
How the representations $R_S$ and $r_S$ are explicitly built from the Young tableau $S$ is a classical topic in representation theory, and we will not need it here. For details we refer the reader to standard textbooks.

What we need to stress here is that a one-to-one correspondence exists between irreducible representations of SU($N$) and $\Sigma_k$. The projector $\mathbb{P}_S$ onto the subspace $V_S$ can be seen as the projector onto the representation $R_S$, or equivalently as the projector onto the representation $r_S$ of $\Sigma_k$. In order to get a useful expression for $\mathbb{P}_S$, we use the Schur orthogonality relation for the representations of the symmetric group:
\begin{equation}
\frac{1}{k!} \sum_{\sigma \in \Sigma_k} r_S(\sigma)^*_{ab} r_{S'}(\sigma)_{a'b'} = \frac{1}{\dim r_S} \delta_{SS'} \delta_{aa'} \delta_{bb'} \ ,
\end{equation}
which can be conveniently written by tracing over the indices $a=b$ as:
\begin{equation}
\frac{\dim r_S}{k!} \sum_{\sigma \in \Sigma_k} \chi_S(\sigma)^* r_{S'}(\sigma) = \delta_{SS'} \mathbf{1}_{S} \ .
\end{equation}
The properly normalized character gives the projector onto the irreducible representation, which is exactly what we are looking for:
\begin{equation}
\mathbb{P}_S = \frac{\dim r_S}{k!} \sum_{\sigma \in \Sigma_k} \chi_S(\sigma)^* r(\sigma) \ ,
\end{equation}
where $\chi_S(\sigma) = \tr \ r_S(\sigma)$ is the character of the representation $r_S$.

We are interested in computing the character $\chi_S(U) = \tr_{R_S}(U)$ of the representation $R_S$.   Taking into account the multiplicity of $R_S$:
\begin{flalign}
& \chi_S(U) = \frac{1}{\dim r_S} \tr [ \mathbb{P}_S R(U) ] \nonumber \\
& = \frac{1}{k!} \sum_{\sigma \in \Sigma_k} \chi_S(\sigma)^* \tr [ r(\sigma) R(U) ] \ .
\label{eq:chimainformula}
\end{flalign}
This formula connects the characters of SU($N$) with the characters of $\Sigma_k$, and is the main ingredient for proving the cancellation of the odd powers of $1/N$ in the sum of RC-conjugate representations of SU($N$). In fact the RC-conjugate representation to $R_S$ is identified by the Young tableau $\widetilde{S}$ obtained from $S$ by swapping rows and columns. The characters of the representation $r_{\widetilde{S}}$ of the symmetric group are related to the characters of $r_S$ by the formula:
\begin{equation}
\chi_{\widetilde{S}}(\sigma) = \textrm{sgn}(\sigma) \chi_S(\sigma) \ ,
\label{eq:RC_char}
\end{equation}
where $\textrm{sgn}(\sigma)$ is the parity of the permutation $\sigma$.\footnote{
The function $\textrm{sgn}$ is a one-dimensional representation of the symmetric group. It corresponds to the Young tableau with a single column. It can be proven that the representations $r_S$ and $r_{\tilde{S}}$ corresponding to two RC-conjugate Young tableaux satisfy~\cite{Stancu:1991rc}:
\begin{equation}
r_{\tilde{S}} = r_S \times \textrm{sgn} \ .
\end{equation}
Since the character of the product of two representations is the product of the characters, Eq.~\eqref{eq:RC_char} follows.
} We recall that a given permutation $\sigma \in \Sigma_k$ can be always decomposed in a product of transpositions (permutations that swap only two elements). The decomposition in transpositions is not unique, however if a decomposition in $m$ transpositions exists then the parity $\textrm{sgn}(\sigma)=(-1)^m$ depends only on the permutation $\sigma$ and not on the particular decomposition.

A useful formula for the character of the representation $R_{\widetilde{S}}$ (RC-conjugate to $R_S$) is obtained by inserting Eq.~\eqref{eq:RC_char} into Eq.~\eqref{eq:chimainformula}:
\begin{equation} \label{eq:RCchimainformula}
\chi_{\widetilde{S}}(U) = \frac{1}{k!} \sum_{\sigma \in \Sigma_k} \textrm{sgn}(\sigma) \chi_S(\sigma)^* \tr [ r(\sigma) R(U) ] \ .
\end{equation}
We will see that the even permutations contribute with even powers of $1/N$ when the expectation value is taken, while the odd permutations contribute with odd powers. Therefore replacing the representation $R_S$ with its RC-conjugate $R_{\widetilde{S}}$ is the same as replacing $N \to -N$.

In order to write the Eqs.~\eqref{eq:chimainformula} and~\eqref{eq:RCchimainformula} more explicitly, we need to recall few facts about permutations. A permutation $\sigma \in \Sigma_k$ can be schematically represented like:
\begin{equation}
\begin{pmatrix}
1 & 2 & \cdots & n \\
\sigma(1) & \sigma(2) & \cdots & \sigma(n)
\end{pmatrix}
\end{equation}
A permutation of $n$ elements is called cyclic (or cycle of length $n$) if it is not possible to split the $n$ elements in two sets that do not mix under the permutation. An example of cyclic permutation is:
\begin{equation}
\begin{pmatrix}
1 & 2 & 3 & 4 \\
4 & 3 & 1 & 2
\end{pmatrix} \equiv ( 1 \ 4 \ 2 \ 3 ) \ ,
\end{equation}
where the notation on the right hand side is a compact way to write that under the cyclic permutation the elements transform as:
\begin{equation}
1 \to 4 \to 2 \to 3 \to 1 \ .
\end{equation}
Given a permutation $\sigma \in \Sigma_k$ of $k$ elements, it is always possible to decompose it in a product of $c(\sigma)$ cycles of length respectively $\ell_1(\sigma) \dots \ell_{c(\sigma)}(\sigma)$. For instance the following permutation:
\begin{equation}
\begin{pmatrix}
1 & 2 & 3 & 4 & 5 & 6 & 7 \\
3 & 4 & 7 & 6 & 1 & 2 & 5
\end{pmatrix} = ( 1 \ 3 \ 7 \ 5 ) \ ( 2 \ 4 \ 6 )
\end{equation}
can be decomposed in two cycles, the first one of length $4$ and the second one of length $3$. Given a cycle of length $\ell$, it can always be decomposed in $\ell-1$ transpositions, as obtained by trivially generalizing the following example:
\begin{equation}
( 1 \ 3 \ 2 \ 5 \ 4 ) = ( 4 \ 5 ) \ ( 5 \ 2 ) \ ( 2 \ 3 ) \ ( 3 \ 1 ) \ .
\end{equation}
The generic permutation $\sigma \in \Sigma_k$ can be hence decomposed in the following number of transpositions:
\begin{equation}
\sum_{q=1}^{c(\sigma)} [\ell_i(\sigma)-1] = k-c(\sigma) \ ,
\end{equation}
and its parity is given by the formula:
\begin{equation} \label{eq:cyclesgn}
\textrm{sgn}(\sigma) = (-1)^{c(\sigma)-k} \ .
\end{equation}
Going back to the traces in Eqs.~\eqref{eq:chimainformula} and~\eqref{eq:RCchimainformula}:
\begin{equation}
\tr [ r(\sigma) R(U) ] = \sum_{(i)} U_{i_{\sigma(1)},i_1} \cdots U_{i_{\sigma(k)},i_k} \ ,
\end{equation}
each cycle in the permutation $\sigma$ closes in a trace with a number of $U$'s equal to the length of the cycle itself:
\begin{equation}
\tr [ r(\sigma) R(U) ] = \prod_{i = 1}^{c(\sigma)} \tr [ U^{\ell_i(\sigma)} ] \ .
\label{eq:tr_rR}
\end{equation}

When we vary $N$ we always keep the representations fixed, in the sense that we keep the corresponding Young tableau fixed. Using the formalism developed above, we want to prove the cancellation of the odd powers of $1/N$ in certain combinations of expectation values of Wilson loops or correlators of Polyakov loops. We will consider the two cases separately.

\textbf{Wilson loops.} If $U$ is a Wilson loop in the pure Yang-Mills theory, then the expectation value of the quantity in Eq.~\eqref{eq:tr_rR} is of order $N$ to the power of the number of traces up to corrections which include only even powers of $1/N$:
\begin{equation}
\langle \tr [ r(\sigma) R(U) ] \rangle = N^{c(\sigma)} f_\sigma \left( \frac{1}{N^2} \right) \ .
\end{equation}
Applying Eq.~\eqref{eq:chimainformula} we get:
\begin{equation}
\frac{1}{N^k} \langle \chi_S(U) \rangle =
\frac{1}{k!} \sum_{\sigma \in \Sigma_k} \chi_S(\sigma)^* N^{c(\sigma)-k} f_\sigma \left( \frac{1}{N^2} \right) \ .
\end{equation}
For the RC-conjugate representation, Eqs.~\eqref{eq:RCchimainformula} and~\eqref{eq:cyclesgn} tell us that a factor $(-1)^{c(\sigma)-k}$ must be inserted in the sum over the permutations:
\begin{equation}
\frac{1}{N^k} \langle \chi_{\widetilde{S}}(U) \rangle =
\frac{1}{k!} \sum_{\sigma \in \Sigma_k} \chi_S(\sigma)^* (-N)^{c(\sigma)-k} f_\sigma \left( \frac{1}{N^2} \right) \ ,
\end{equation}
which is equivalent to change $N \to -N$. This argument explicitly shows that the combination
\begin{equation}
\frac{\langle \chi_S(U) \rangle + \langle \chi_{\widetilde{S}}(U) \rangle }{N^k}
\end{equation}
contains only even powers of $1/N$. A trivial consequence of this
proof is that for an RC-selfconjugate representation $M$ with $k$ sources
$\langle \chi_M(U) \rangle/N^k $ contains only even powers of $1/N$.

\textbf{Polyakov loops.} If $\Omega$ and $\Omega'$ are two Polyakov loops in the pure Yang-Mills theory, their correlator can be written using Eq.~\eqref{eq:chimainformula} as:
\begin{flalign}
& \langle \chi_S(\Omega)^\dag \chi_S(\Omega') \rangle = \nonumber \\
& = \frac{1}{(k!)^2} \sum_{\sigma,\sigma' \in \Sigma_k}
\chi_S(\sigma) \chi_S(\sigma')^* \times \nonumber \\
& \quad \times \langle \tr [ r(\sigma) R(\Omega) ]^\dag \tr [ r(\sigma') R(\Omega') ] \rangle \ .
\end{flalign}
Using again Eq.~\eqref{eq:tr_rR} one can count the number of traces in the expectation values in the r.h.s. of the last equation:
\begin{flalign}
& \langle \tr [ r(\sigma) R(\Omega) ]^\dag \tr [ r(\sigma') R(\Omega') ] \rangle \nonumber \\
& =
N^{c(\sigma)+c(\sigma')} g_{\sigma,\sigma'} \left( \frac{1}{N^2} \right) \ .
\end{flalign}
Because of the center symmetry, it can happen that the function $g_{\sigma,\sigma'}$ goes to zero in the large-$N$ limit, or even that it is identically zero. It is not important to distinguish those cases at this stage. Putting all together we get:
\begin{flalign}
& \langle \chi_S(\Omega)^\dag \chi_S(\Omega') \rangle = \nonumber \\
& \frac{1}{(k!)^2} \sum_{\sigma,\sigma' \in \Sigma_k}
\chi_S(\sigma) \chi_S(\sigma')^*
N^{c(\sigma)+c(\sigma')} g_{\sigma,\sigma'} \left( \frac{1}{N^2} \right) \ .
\end{flalign}
For the RC-conjugate representation, Eqs.~\eqref{eq:RCchimainformula} and~\eqref{eq:cyclesgn} tell us that a factor $(-1)^{c(\sigma)+c(\sigma')-2k} = (-1)^{c(\sigma)+c(\sigma')}$ must be inserted in the sum over the permutations:
\begin{flalign}
& \langle \chi_{\widetilde{S}}(\Omega)^\dag \chi_{\widetilde{S}}(\Omega') \rangle = \nonumber \\
& = \frac{1}{(k!)^2} \sum_{\sigma,\sigma' \in \Sigma_k}
\chi_S(\sigma) \chi_S(\sigma')^*
(-N)^{c(\sigma)+c(\sigma')} g_{\sigma,\sigma'} \left( \frac{1}{N^2} \right) \ ,
\end{flalign}
which is equivalent to change $N \to -N$. This argument explicitly shows that the combination
\begin{equation}
\langle \chi_S(\Omega)^\dag \chi_S(\Omega') \rangle + \langle \chi_{\widetilde{S}}(\Omega)^\dag \chi_{\widetilde{S}}(\Omega') \rangle
\end{equation}
contains only even powers of $1/N$. A trivial consequence of this
proof is that for an RC-selfconjugate representation $M$ with $k$ sources
$ \langle \chi_M(\Omega)^\dag \chi_M(\Omega') \rangle $ contains only even powers of $1/N$.

\begin{figure}[t!]
\begin{center}
\includegraphics[height=3.0cm]{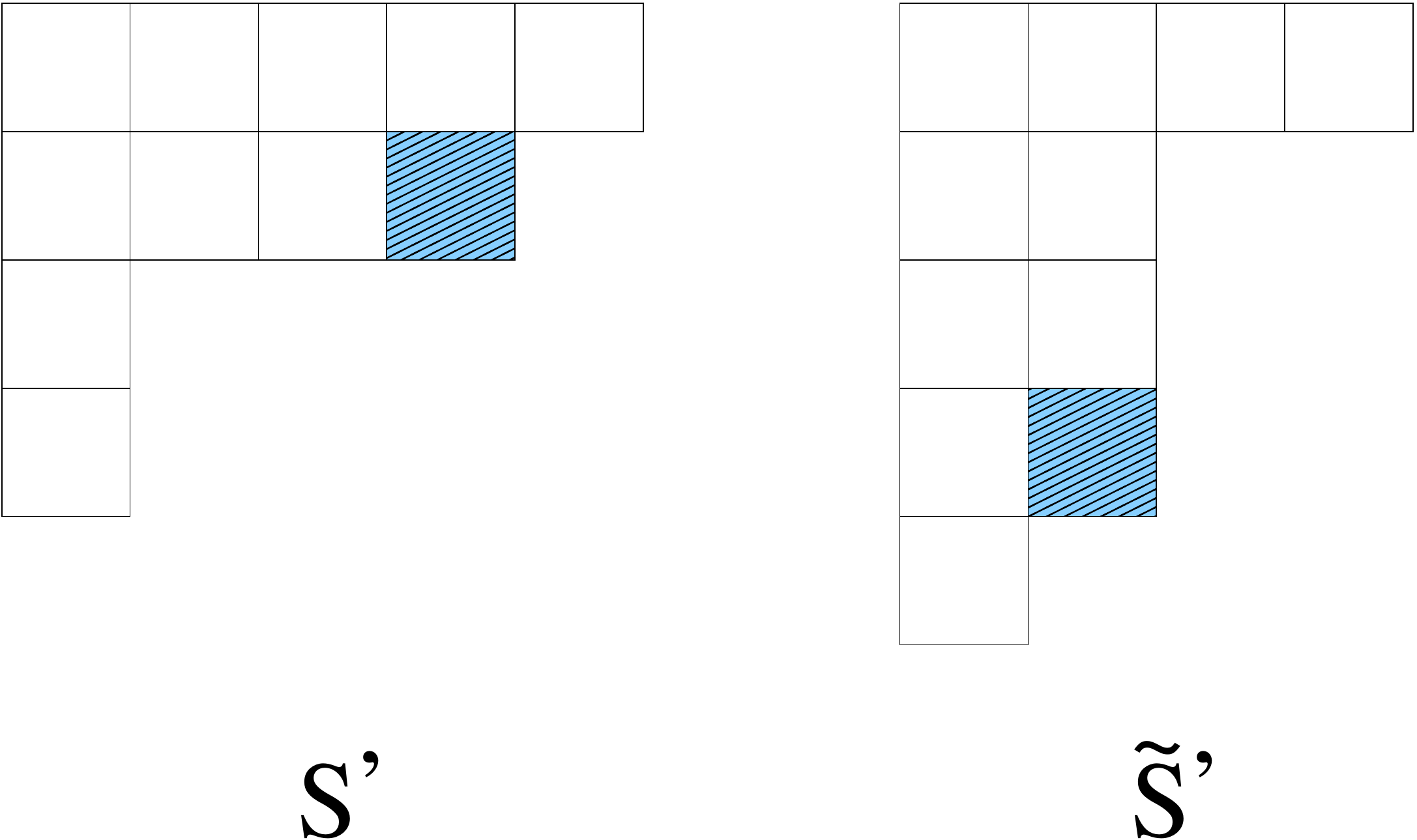}    
\end{center}
\caption{The RC-conjugate pair of Young tableaux $(S^{\prime},\widetilde{S}^{\prime})$ obtained from
  $(S,\widetilde{S})$ (see Fig.~\ref{fig:young_conj}) adding the shaded box in
  conjugate positions.}
\label{fig:young_conj_2}
\end{figure}

\section{Casimirs and dimensions in RC-conjugate representations}
\label{appendix:casimirs}
In this Appendix, we shall prove
Eqs.~\eqref{eq:conjcasimirs_0}~and~\eqref{eq:conjdimensions}. 

The dimension of a representation $R_S$ associated with the Young tableau $S$ is given by the ratio
\beq
d = A/D \ ,
\eeq
where the numerator $A$ is obtained from the tableau by labeling the box in row $i$
and column $j$ with $N + j - i$ (the rows are
numbered from top to bottom and the columns from left to right; both
indices start from one) and multiplying all labels together (see e.g.~\cite{Cheng:1985bj}). The
denominator $D$ is obtained by labeling each box with the total
number of boxes to the right and to the bottom of it. The conjugation
of the tableau leaves $D$ invariant, since it interchanges right
with bottom, and sends $j-i$ into $i - j$, since it interchanges rows
with columns. If we consider $A/N^k$, with $k$ the number of boxes
(assumed to be less than $N/2$), this is the product of factors of the form
\beq
f_{ij} = (1 + (j-i)/N)\ .
\eeq
After conjugation, each term in $1/N$ changes sign. Together with the
invariance of $D$, this implies that even powers of $1/N$ in $d$ stay
invariant after conjugation, while odd powers of $1/N$ change
sign. This proves Eq.~\eqref{eq:conjdimensions}. For an RC-selfconjugate
tableau, for each term of the form $f_{ij}$ there is a term
$f_{ji}$. Due to the antisymmetry of $f_{ij}$ under the exchange of
$i$ and $j$, this implies that dimensions of representations with RC-selfconjugate
tableaux do not contain odd powers of $N$.

As for Eq.~\eqref{eq:conjcasimirs_0}, using the fact that $C_F(N) = -
C_F(-N)$, we notice that this statement is equivalent to 
\beq
\label{eq:relcasimir}
C_{\widetilde{R}}(N)  = - C_R(-N) 
\eeq
for the representations corresponding to the two RC-conjugate Young
tableaux $S$ and $\widetilde{S}$. Eq.~\eqref{eq:relcasimir} is true for the
two-index symmetric and antisymmetric representations, which have
RC-conjugate Young tableaux. We prove~\eqref{eq:relcasimir}
recursively. Young tableaux for RC-conjugate representations of rank $k$ can be generated
from Young tableaux of RC-conjugate representations of rank $k-1$ by
adding a box in conjugate positions. For instance,
Fig.~\ref{fig:young_conj_2} shows the tableaux $S^\prime$ and
$\widetilde{S}^\prime$ obtained from $S$ and $\widetilde{S}$ given in
Fig.~\ref{fig:young_conj}. Let us assume we have proved
Eq.~\eqref{eq:relcasimir}  for all representations $R$ and $\widetilde{R}$ of rank $k-1$. We will
show that this implies that Eq.~\eqref{eq:relcasimir} is true also
for all $R^{\prime}$ and $\widetilde{R}^{\prime}$ of rank $k$. Given the RC-conjugate
pair ($R^{\prime}$,$\widetilde{R}^{\prime}$) of rank $k$, an RC-conjugate pair ($R$,$\widetilde{R}$) of
rank $k-1$ exists such that ($S^{\prime}$,$\widetilde{S}^{\prime}$) can be
obtained adding an extra box in conjugate positions. From
e.g. appendix A.3 of~\cite{Lucini:2001nv}, we get 
\beq
C_{R^{\prime}}(N) = C_R(N) + \frac{N}{2} + (n_m - m) - \frac{2k - 1}{2N} \ ,
\eeq
$n_m$ being the number of boxes in row $m$ (the only row for which the
Young tableaux $S$ and $S^{\prime}$ have a different number of boxes)
of $S^{\prime}$. Similarly, 
\beq
C_{\widetilde{R}^{\prime}}(N) = C_{\widetilde{R}}(N) + \frac{N}{2} + (n_l - l) - \frac{2k - 1}{2N} \ .
\eeq
Now, since the extra boxes with respect to $(S,\widetilde{S})$ are in conjugate
positions, $n_l = m$ and $l = n_m$, which yields to
\beq
C_{\widetilde{R}^{\prime}}(N) = C_{\widetilde{R}}(N) + \frac{N}{2} + (m - n_m) - \frac{2k - 1}{2N} \ .
\eeq
Hence
\beq
C_{\widetilde{R}^{\prime}}(N) &=& \phantom{-} C_{\widetilde{R}}(N) + \frac{N}{2} + (m - n_m)- \frac{2k - 1}{2N} \\ \nonumber \\
                             &=& - C_R(-N) + \frac{N}{2} - (n_m - m) - \frac{2k - 1}{2N} \nonumber \\
                             &=& -C_{R^{\prime}}(-N) \ ,
\eeq
which is the statement we needed to prove. A consequence of this
property is that in Casimirs of representations with RC-selfconjugate
tableaux only odd powers of $N$ appear. As a result, if $M$ is a
representation with an RC-selfconjugate tableaux, the ratio $C_M/C_F$
contains only even powers of $N$.

\bibliography{ksprd}

\end{document}